\documentclass[sigconf,edbt]{acmart-edbt2020}

\def\BibTeX{{\rm B\kern-.05em{\sc i\kern-.025em b}\kern-.08em
    T\kern-.1667em\lower.7ex\hbox{E}\kern-.125emX}}

\usepackage{mathtools}
\usepackage{amsmath,amsthm, amsfonts}
\usepackage{dsfont}
\usepackage{graphicx}
\usepackage{balance}  
\usepackage{booktabs}
\usepackage{caption}
\usepackage{url}
\usepackage{fixmetodonotes}
\usepackage{subcaption}
\usepackage{bm}
\usepackage{subcaption}

\usepackage{booktabs} 

\renewcommand\footnotetextcopyrightpermission[1]{} 

\begin{document}
\title{RETRO: Relation Retrofitting For In-Database Machine Learning on Textual Data}

\author{Michael G\"unther}
\affiliation{%
  \institution{Database Systems Group, Technische Universit\"at Dresden}
  \streetaddress{Nöthnitzer Str. 46}
  \city{Dresden}
  \country{Germany}
  \postcode{D-01062}
}
\email{Michael.Guenther@tu-dresden.de}

\author{Maik Thiele}
\affiliation{%
  \institution{Database Systems Group, Technische Universit\"at Dresden}
  \streetaddress{Nöthnitzer Str. 46}
  \city{Dresden}
  \country{Germany}
  \postcode{D-01062}
}
\email{Maik.Thiele@tu-dresden.de}

\author{Wolfgang Lehner}
\affiliation{%
  \institution{Database Systems Group, Technische Universit\"at Dresden}
  \streetaddress{Nöthnitzer Str. 46}
  \city{Dresden}
  \country{Germany}
  \postcode{D-01062}
}
\email{Wolfgang.Lehner@tu-dresden.de}

\renewcommand{\shortauthors}{}

\begin{abstract}
There are massive amounts of textual data residing in data\-bases, valuable for many machine learning (ML) tasks.
Since ML techniques depend on numerical input representations, word embeddings are increasingly utilized to convert symbolic representations such as text into meaningful numbers.
However, a na\"ive one-to-one mapping of each word in a database to a word embedding vector is not sufficient and would lead to poor accuracies in ML tasks.
Thus, we argue to additionally incorporate the information given by the database schema into the embedding, e.g. which words appear in the same column or are related to each other.
In this paper, we therefore propose \emph{RETRO} (RElational reTROfitting), a novel approach to learn numerical representations of text values in databases, capturing the best of both worlds, the rich information encoded by word embeddings and the relational information encoded by database tables.
We formulate \textit{relation retrofitting} as a learning problem and present an efficient algorithm solving it.
We investigate the impact of various hyperparameters on the learning problem and derive good settings for all of them.
Our evaluation shows that the proposed embeddings are ready-to-use for many ML tasks such as classification and regression and even outperform state-of-the-art techniques in integration tasks such as null value imputation and link prediction.
\end{abstract}

%
%


\setcopyright{none}
\maketitle

\section{Introduction}
Relational databases reflect the most common infrastructure to store large amounts of structured and unstructured data persistently.
According to a recent survey conducted by Kaggle (2017), relational data is reported as the most popular data model by 14,000 data scientists, with at least 65\% working daily with it.
Consequently, industry and academia built a plethora of in-databases ML systems such as MADlib, MLlib, SimSQL, SQLML, SciDB, and SAP PAL.
However, there is very little work on automatically deriving high-quality numerical representations of textual data residing in databases without any manual effort.
\begin{figure}[t]
 \includegraphics[width=0.47\textwidth]{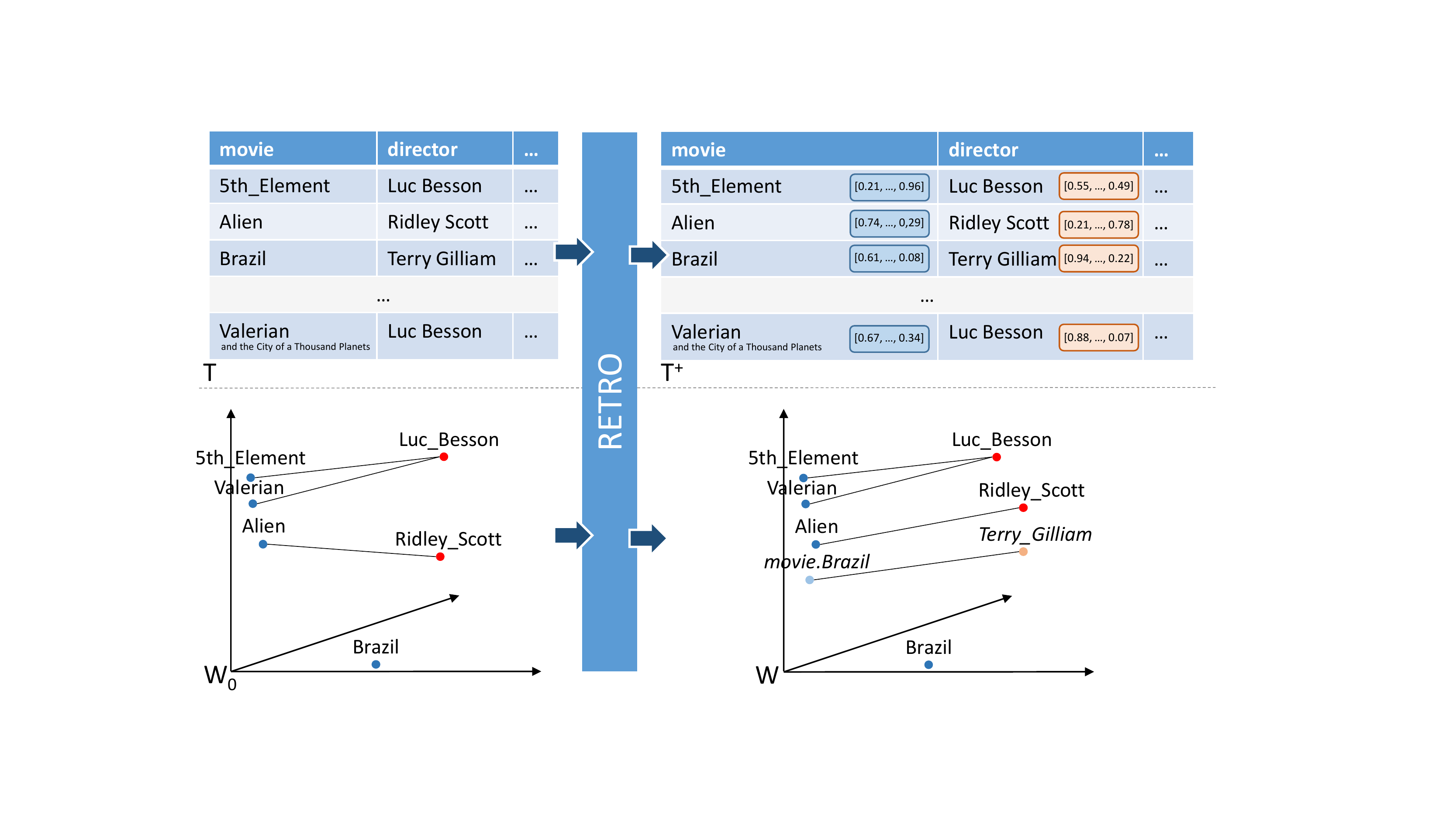}
 \caption{Relational Retrofitting: original word embedding $W^0$ and relation $T$ (left), retrofitted word embedding $W$ and augmented relation $T^{+
}$ (right)}
 \label{fig:intro_example}
\end{figure}
In this paper, we are investigating a learned vector representation for database tables, ready-to-use for a wide range of ML, retrieval, and data cleaning tasks such as classification, regression, null value imputation, entity resolution, and many more.
Specifically, we extend the concept of re\-tro\-fit\-ting \cite{faruqui2014retrofitting} to inject the knowledge encoded by a table $T$ into a word embedding $W^0$ to find good descriptors for all elements of $T$. The example sketched in Figure~\ref{fig:intro_example} highlights the main challenges involved with this approach.
In this example, a movie table $T$ is retrofitted into an existing word embedding $W^0$.
Given the movie table, it is known that all entities within the movie column must be movies, although some of the titles, such as ``Brazil" or ``Alien", may be interpreted in different ways.
Word embeddings themselves do not provide an accurate representation of text values since their meaning could be ambiguous in general, e.g. ``Brazil" that has a larger distance to the movie titles, while they have specific meanings within a database model.
In the same sense, $T$ provides a specific movie-director relationship.
However, the relationship between two entities in word embeddings can be much richer, e.g. if the director of a movie is also the producer or one of the actors.
Finally, the set of terms appearing in $T$ and $W^0$ are often disjunct.
Since the vocabulary of a word embedding is limited due to a frequency threshold, many terms appearing in a database will have no counter-part within the embedding.\\
We present \emph{RETRO}, a novel relational retrofitting approach (depicted in Figure~\ref{fig:intro_example}), addressing all these challenges.
\emph{RETRO} augments all terms in database tables by dense vector representations encoding the semantics given by the table and the word embedding. The advantages of \emph{RETRO} are manifold:
\begin{itemize}
	\item \emph{RETRO} exploits the knowledge encoded by pre-trained word embeddings to find meaningful vectors for all terms in a database.
	\item \emph{RETRO} is able to reflect relational meanings of words that cannot be captured by pure word co-occurrences \cite{lenci2018} and modify word vectors to specialize in a specific task.
	\item \emph{RETRO} does not rely on re-training, which allows us to incrementally maintain the word vectors whenever the data in the database changes. Hence, the learned representations are incrementally updateable which is an important requirement for in-database ML systems.
	\item \emph{RETRO} is applicable to different types of word vectors, either pre-trained or specifically trained for a given domain.
\end{itemize}
\pagebreak
Besides the aforementioned ML tasks, our relational vector representations can be also applied within information retrieval, data integration, and database exploration techniques such as keyword search on relational data, table augmentation, link prediction, table clustering, table summarization, and many more.
Finally, \emph{RETRO} may be used for domain-specific similarity queries as proposed in~\cite{Bordawekar:2017:UWE:3076246.3076251, gunther2018freddy}.\\\\
\textbf{Outline and Contribution.} The remainder of the paper is organized as follows:
In Section~\ref{sec:problemscope}, we give an overview of the problem and a brief walkthrough of our approach.
The details of data pre-processing are given in Section~\ref{sec:extraction} including the initialization of word vectors to train on and the extraction of the relational information out of the input tables.
We then present our novel relational retrofitting approach \emph{RETRO} in Section~\ref{sec:retrofitting}.
In detail, we formulate \textit{relation retrofitting} as a learning problem with a word embedding and a set of relational tables as input.
We investigate the impact of various hyperparameters on the learning problem and derive good settings for all parameters.
We further prove the convexity of its underlying objective function.
Given this property the learning converges to an optimal representation and does not depend on the initialization.
Additionally, we present different optimizations which simplify the optimization function under certain assumptions, leading to lower runtimes.
To demonstrate the potential of \emph{RETRO} in real-world scenarios, we evaluate the performance of \emph{RETRO} within different classification and regression tasks in Section~\ref{sec:eval}.
Here\-by, we compare our method to node embedding techniques \cite{perozzi2014deepwalk} and the original proposed retrofitting approach~\cite{faruqui2014retrofitting}.
We show the feasibility of \emph{RETRO} in automatically creating vector representations for database tables, outperforming state-of-the-art techniques in many downstream tasks.
Finally, we present related work in Section~\ref{sec:relatedwork} and conclude in Section~\ref{sec:conclusion}.
%
\section{Problem Scope}
\label{sec:problemscope}
Due to their appealing properties, word embedding techniques such as word2vec~\cite{NIPS2013_5021}, GloVe~\cite{pennington2014glove} or fastText~\cite{bojanowski2017enriching} have become conventional wisdom in many research fields such as NLP or information retrieval to represent text values.
Word vectors are trained by processing large text corpora.
For example, popular pre-trained GloVe embedding set\footnote{\url{https://nlp.stanford.edu/projects/glove/}} was trained on a corpus with 840 billion tokens.
The resulting dense vectors are able to capture word similarities and relations such as word analogies, gender-inflections, or geographical relationships.\\
We aim at leveraging these semantically rich embeddings to generate good vector representations for text values residing within relational databases.
We extend therefore the notion of retrofitting which was initially proposed in \cite{faruqui2014retrofitting} to incorporate knowledge from a word similarity graph into word embeddings.
Instead of using relational information directly during the training objective of word vectors, retrofitting is performed as a post-processing step on the word embeddings.
This allows to apply retrofitting to different vector models trained with different methods (e.g. word2vec~\cite{NIPS2013_5021}, GloVe~\cite{pennington2014glove} or fastText~\cite{bojanowski2017enriching})\\
However, while retrofitting is typically used to improve the vector quality of general-purpose word embeddings for certain domains and tasks, we aim at augmenting each text entry in database tables with word vectors 1) reflecting the semantics of the relation and 2) fitting into the given embedding.
\begin{figure*}[t]
 \centering
 \includegraphics[width=\linewidth]{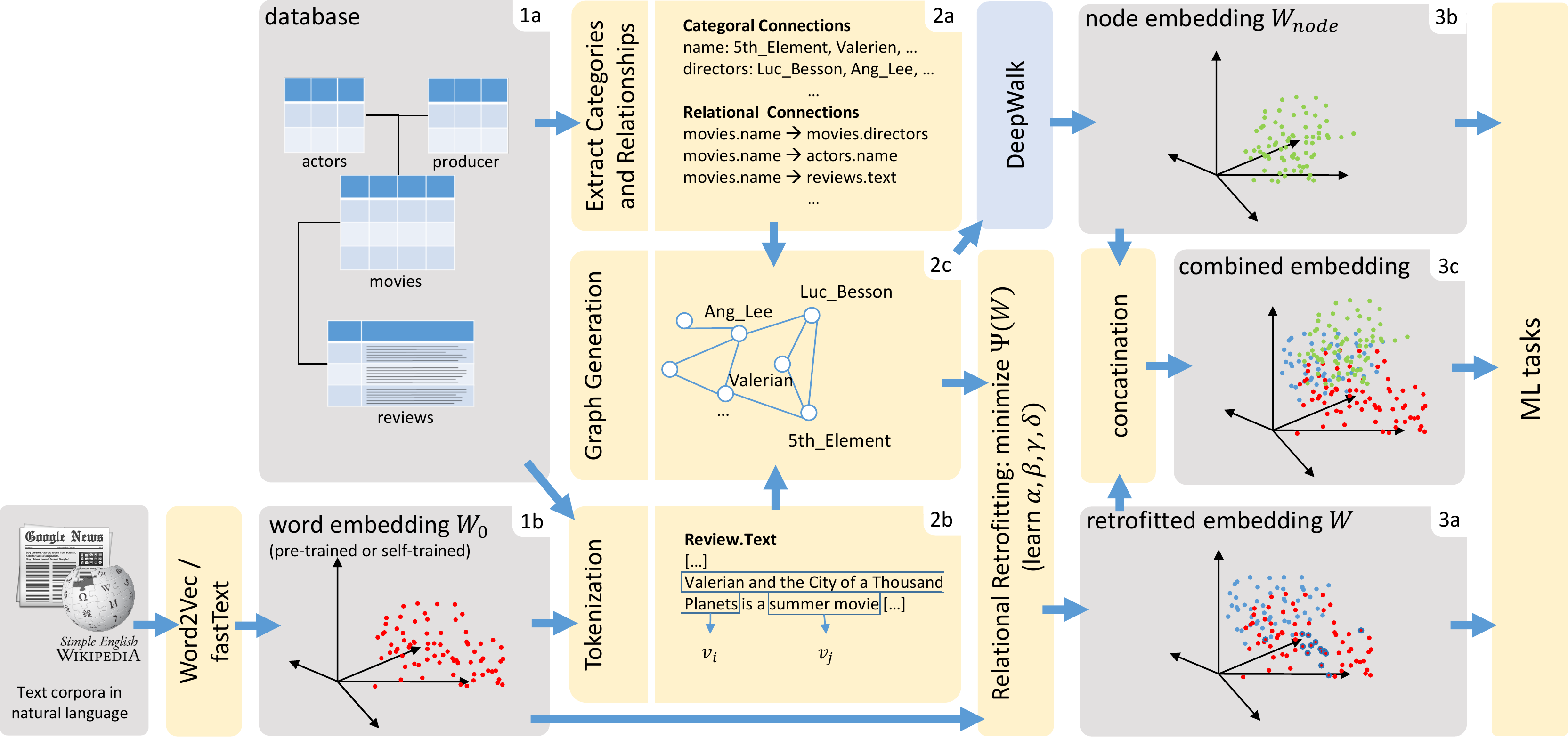}
 \caption{Overview of the Relational Retrofitting Process}
 \label{fig:embedding-process}
\end{figure*}
An overview of our \textit{relational re\-tro\-fit\-ting} approach is depicted in Figure~\ref{fig:embedding-process}.
The input of the overall process consists of a database (Step 1a) as well as a given word embedding $W^0$, either pre-trained or self-trained for a specific domain (Step 1b).
In a first step (Step 2a), all database text values are extracted together with the information in which column they appeared, capturing their ``category".
For the given example, entries in the movie column belong to the category \emph{movie}.
Likewise, the attribute \emph{director} forms another category.
Further the relationships between all text values are extracted, e.g. relationships between \textit{movie} and \emph{director} as well as primary key foreign key relations between \emph{movie} and \emph{actors}, \emph{producers} and \emph{reviews}.\\
Multi-word phrases for which word embeddings are available are preserved by a specific tokenization approach (Section~\ref{sec:tokenization}).
Words in the database having no counter-part in the given embedding $W^0$ are initialized with a null vector.
The extracted relationships (Step 2a) and the tokenized text values (Step 2b) are combined to a property graph representation (Step 2c).
The graph encodes all the relations between text values according to the given database schema.
Our relational re\-tro\-fit\-ting approach \emph{RETRO} (Section~\ref{sec:retrofitting}) adapts the base word embedding representation $W^0$ using the set of relationships encoded in the graph.
The result will be the retrofitted embeddings (Step 3a) containing vectors for all terms appearing in the input database.\\
In addition to the core algorithm, we investigate an existing node embedding technique called DeepWalk~\cite{perozzi2014deepwalk} taking the already derived graph representation (Step 2c) as an input (see Section~\ref{sec:nodeembedding}).
Node embedding approaches can encode relational information more accurately than retrofitting approaches which also have to retain the connection to the word representation.
Hence, node embeddings (Step 3b) should perform better than retrofitted embeddings (Step 3a) when relational features are prevalent in an ML task in opposite to pure to textual information.
Therefore, we investigate how node embeddings can be trained on relational databases and combined with retrofitted embeddings (Step 3c).
\section{Preprocessing of Textual and Relational Data}
\label{sec:extraction}
Our relational retrofitting approach learns a matrix of vector representations $W = (\bm{v_1},\ldots \bm{v_n})$ with $\bm{v_i} \in \mathbb{R}^D$ for every unique text value $T= (\bm{t_1},\ldots \bm{t_n})$ in a database.
To find initial word vectors for every text value we use a tokenization approach presented in Section~\ref{sec:tokenization}
These vectors are stored in a matrix $W^0 = (\bm{v_1'},\ldots \bm{v_n'})$ forming the basis for the retrofitting process.
In addition, \emph{categorial and relational connections} are extracted from the database (see Section~\ref{sec:db_rel_ext}).
This encompasses semantic relations between text values, which are derived from the relational schema.
Those connections are used later by \emph{RETRO} to create a representation capturing the context of the text value in the database and thus helps to preserve their semantic more accurately compared to a plain word embedding representation.
In order to apply DeepWalk, all information is condensed into a common graph representation shortly presented in Section~\ref{sec:graphgeneration}.
\subsection{Out-of-Vocabulary Problem and Tokenization}
\label{sec:tokenization}
In practice, we observe many database text values $t_i \in T$ have no corresponding vector $\bm{v_i}$ in the word embedding and thus can not directly be represented as word vectors.
This is especially true for named entities which occur very often in database tables but also for longer text value like reviews which consist of multiple words.
The so-called \emph{out of vocabulary (OOV)} problem emerges from the limitation that embeddings are present only for words and phrases occurring frequently in the training corpus.
A very simple method is to run a standard tokenization approach, collect the corresponding embeddings for every token and calculate the centroid of the collected word vectors.
However, multi-token-words are often tokenized in a wrong way leading to bad word vector assignments.
For example, there is a specific embedding for ``bank account'', however, simple tokenization would assign the words to the separate embeddings ``bank'' and ``account''.
Large word embedding datasets consist mainly out of multi-words and phrases which are ignored in this way.
To provide a good initial vector for each text value, we propose the following tokenization approach:
First, a lookup trie (prefix tree) is created for the dictionary of the given word embedding dataset, where every node represents a token.
By considering the lookup trie the longest possible sequences of nodes are extracted (e.g. ``bank account'' instead of ``bank''), resulting in a bag of tokens for each text value.
The initial embedding of a text value is determined by averaging the corresponding vectors of it's tokens.
Averaging or similar combination techniques for word vectors are frequently used to represent phrases~\cite{alghunaim2015vector, zhou2015representation}.
Text values containing no token with a corresponding embedding, initially get assigned a null vector.
However, since \emph{RETRO} is also considering the relations to other text values, these null vectors get assigned a meaningful representation in the learning phase.
\subsection{Relationship Extraction}
\label{sec:db_rel_ext}
We extract the column information or rather \emph{category} of all text values and their \emph{relationships} to all other text values.
Therefore, all necessary schema information and the actual data is queried from the database.\\\\
\textbf{Categorial Connections:} For every text column in the data\-base we define a category $C \in \{C_1,\ldots,C_m\}$ containing all text values in the column.
Respectively, the category of an embedding $\bm{v_i} = W_{i*}$ is referenced by $C(i)$.\\\\
\textbf{Relational Connections:} We divide relationships into multiple relation groups ref\-er\-enced by sets $E_r \subseteq \{(i,j) | i,j \in \{1,\ldots,n\}\}$ where $r \in R$ can be considered as a label (e.g. 'movie$\to$director').
Every relation group contains relations of a specific type (e.g. connects movies to their directors).
The elements of a relation group are pairs of indices of the word vectors which correspond to row ids in the matrix $W$.
There are different kinds of relationships in the database from which a relation group could be extracted:\\
a) A row-wise relationship is formed by two text columns $A$ and $B$ appearing in the same table, i.e. it connects $t_i \in A$ and $t_j \in B$ whereas $t_i$ appears in column $A$ and $t_j$ appears in column $B$ in the same row.\\
b) A primary key-foreign key relationship (one-to-many) connects a text value $t_i$ in a specific column from one table with a text value $t_j$ from a column in another table.\\
c) Many-to-many relationships are formed by two text columns related by a link table with foreign key pairs.\\
Each text column pair connected by the aforementioned relationships forms a relation group $E_r$.
The inverted counterpart of this relation group denoted by $E_{\bar{r}} = \{(j,i) | (i,j) \in E_r)\}$.
\subsection{Uniqueness of Text Values}
In the process of generating word vectors for every unique text value in a database we have to distinguish two cases:\\
1) If the same text value occurs $n$ times in different columns, i.e. within different categories, $n$ separate embeddings are created.
In the context of our movie example, the word ``Am\'elie'' may occur as a person and as a movie title leading to two distinct embeddings (in $W$).\\
2) If the same text value occurs $n$ times in the same column, i.e. the same category, only a single embedding is created.
Even though different people could have the name ``Am\'elie'' and can be distinguished by different keys there is only a single embedding.\\
While in the second case it might be plausible to create multiple embeddings, we argue that text values are often used for categorial types like \emph{male} and \emph{female} where a separate embedding for each occurrence would be unfavorable since it creates a high amount of embeddings referring to the same concept.
\subsection{Graph Generation}
\label{sec:graphgeneration}
A graph representation is a prerequisite for training node embeddings such as DeepWalk.
To compile a graph $G=(V,E)$ the text values extracted by the tokenization process (Section~\ref{sec:tokenization}) together with categorial and relational connections (Section~\ref{sec:db_rel_ext}) are combined.
The node-set $V=V_C\cup V_T$ consists of text value nodes $V_T$ for every distinct text value in a database column and blank nodes for every column $V_C$.
The edge set $E=\bigcup_{r \in R} E_r\cup E_C$ consists of a set of edges $E_r$ for every relational type and edges $E_C$ connecting text values of one column to a common category node.
\section{RETRO: Methodology}
\label{sec:retrofitting}
Before introducing a formal notion of the retro\-fitting approach in Section~\ref{sec:relationalretrofittingproblem}, we briefly review the original re\-tro\-fit\-ting problem proposed by Faruqui et al. in Section~\ref{sec:faruqui}.
In Section~\ref{sec:retrofitting_algorithm}, we present an algorithm using matrix operations solving the relation retrofitting efficiently.
Favourable settings for the hyperparameters of our proposed algorithm are discussed in Section~\ref{subsec:paramaterconf}.
In Section~\ref{sec:optimization} we additionally outline potential optimizations.
\subsection{Original Retrofitting Problem}
\label{sec:faruqui}
Faruqui et al. first proposed a retrofitting approach taking word embeddings in a matrix $W^0$ and a graph $G = (Q, E_F)$ representing a lexicon as input \cite{faruqui2014retrofitting}.
The retrofitting problem was formulated as a dual objective optimization function:
The embeddings of the matrix $W$ are adapted by placing similar words connected in the graph $G$ closely together, while at the same time the neighborhood of the words from the original matrix $W^0$ should be preserved.
Hereby, $Q = \{q_1, \ldots, q_n\}$ is a set of nodes where each node $q_i$ corresponds to a word vector $\bm{v_i}\in W^0$ and $E_F \subset \{(i,j) | i, j \in \{1,\ldots,n\}\}$ is a set of edges.
The graph is undirected, thus $(i,j) \in E_F \Leftrightarrow (j,i) \in E_F$.
The authors specified the retrofitting problem as a minimization problem of the following loss function:
\begin{flalign}\label{eq:opt_origin}
\Psi_\mathit{Faruqui}(W) = \sum_{i=1}^n \Big[\alpha_i||\bm{v_i}-\bm{v_i'}||^2 + \sum_{j:(i,j)\in E_F}\beta_i||\bm{v_i}-\bm{v_j}||^2\Big]
\end{flalign}
The constants $\alpha_i$ and $\beta_i$ are hyperparameters.
$\Psi_\mathit{Faruqui}(W)$ is convex for positive values of $\alpha_i$ and $\beta_i$.
Thus, the optimization problem can be solved by an algorithm, which iterates over every node in $Q$ and updates the respective vector in $W$ according to Equation \eqref{eq:d-update}.
The update function for any vector $\bm{v_i}$ is obtained from the root of the partial derivative $\frac{\partial\Psi_\mathit{Faruqui}(W)}{\partial\bm{v_i}}$.
\begin{flalign}
	\label{eq:d-update}
	\bm{v_i} = \frac{\alpha_i \bm{v_i'} + \sum\limits_{\mathclap{j:(i,j)\in E_F}}(\beta_i+\beta_j) \bm{v_j}}{\alpha_i+\sum\limits_{\mathclap{j:(i,j)\in E_F}}(\beta_i+\beta_j)}
\end{flalign}
However, Faruqui et al. decided to use this simplified update function:
\begin{flalign}
	\label{eq:mf-simple-update}
	\bm{v_i} = \frac{\alpha_i \bm{v_i'} + \sum\limits_{\mathclap{j:(i,j)\in E_F}}\beta_i \bm{v_j}}{\alpha_i+\sum\limits_{\mathclap{j:(i,j)\in E_F}}\beta_i}
\end{flalign}
While there is no justification in the paper, updates according to Equation~\eqref{eq:mf-simple-update} get along with some advantage.
The update function in Equation~\eqref{eq:d-update} can shift embeddings strongly in order to prevent an adaptation of its neighboring nodes.
This leads to a higher loss in the summand of this single vector but lowers the value of the global loss function.
Since the result of a classification task rather depends on a few embeddings and not on the whole embedding corpus, this might not be advantageous.
The calculation using updates as in Equation~\eqref{eq:mf-simple-update} does not solve the optimization problem in Equation~\eqref{eq:opt_origin}, however, can be considered as a formulation of a recursively defined series.
For every $\bm{v_i} \in W$ the series converges to a specific vector which is the desired output of the retrofitting process.
\subsection{Relational Retrofitting Problem}
\label{sec:relationalretrofittingproblem}
\emph{RETRO} considers relational and categorial connections (see Section~\ref{sec:db_rel_ext}) to retrofit an initial embedding.
Accordingly, we define a loss function $\Psi$ apdapting embeddings to be similar to their original word embedding representation $W^0$, the embeddings appearing in the same column, and related embeddings.
\begin{flalign}
\Psi(W) = \sum_{i = 1}^n \Big[ \alpha_i ||\bm{v_i} - \bm{v_i'}||^2 + \beta_i \Psi_C(\bm{v_i}, W) + \Psi_R(\bm{v_i}, W) \Big]
\end{flalign}
The categorial loss is defined by $\Psi_C$ and treats every embedding $\bm{v_i}$ to be similar to the constant centroid $\bm{c_i}$ of the original embeddings of text values in the same column $C(i)$.
\begin{flalign}
\Psi_C(\bm{v_i}, W) = ||\bm{v_i} - \bm{c_i}||^2 \;\; \bm{c_i} = \frac{\sum\limits_{j \in C(i)}{\bm{v_j'}}}{|C(i)|}
\end{flalign}
The relational loss $\Psi_R$ treats embeddings $v_i$ and $v_j$ to be similar if there exists a relation between them and dissimilar otherwise.
$E_r$ is the set of tuples where a relation $r$ exists.
$\widetilde{E_r}$ is the set of all tuples $(i,j)\not\in E_r$ where $i$ and $j$ are part of relation $r$.
Thus, each of both indices has to occur at least in one tuple of $E_r$.
\begin{flalign}
&\Psi_R(\bm{v_i}, W) &=& \sum_{r \in R} \Big[ \sum_{\mathclap{\substack{j:(i,j) \\ \in E_r}}} \gamma^r_{i} || \bm{v_i} - \bm{v_j} ||^2 - \sum_{\mathclap{\substack{k:(i,k) \\ \in \widetilde{E_r}}}} \delta^r_{i}|| \bm{v_i} - \bm{v_k} ||^2 \Big] && 
\end{flalign}
$\alpha$, $\beta$, $\gamma$ and $\delta$ are hyperparameters.
$\Psi$ has to be a convex function.
In the appendix we prove that convexity is assured if the hyperparameters fulfill the following inequalities:
\begin{flalign}\label{eq:convex_condition}
& \forall r \in R, i \in \{1,\ldots,n\}\;\;( \alpha_i  \geq 0,\;\; \beta_i  \geq 0,\;\; \gamma_i^r \geq 0 )\\
\nonumber
& \forall \bm{v_i}\in W \;\;( 4\alpha_i - \sum_{r\in R}\sum_{j:(i,j)\in\widetilde{E_r}}\delta_i^r \geq 0)
\end{flalign}
In practice, however, other parameter configurations which do not comply might work as well.
Given the property of convexity, an iterative algorithm can be used to minimize $\Psi$.
This algorithm iteratively executes for all $\bm{v_i} \in V$ the following equation, which is derived from the root of the partial derivative $\frac{\partial\Psi(W)}{\partial\bm{v_i}}$.
\begin{flalign}\label{eq:iterative1}
  \bm{v_i} =& \frac{\alpha_i\bm{v_i'} + \beta_i\bm{c_i} +  \sum_{r \in R}\Big[ \sum_{\mathclap{\substack{\vspace{2mm}\\ j:(i,j) \\ \in E_r}}} (\gamma^r_i+\gamma^{\bar{r}}_j) \bm{v_j} - \sum_{\mathclap{\substack{\vspace{2mm}\\k:(i,k) \\ \in \widetilde{E_r}}}} (\delta^r_i+\delta^{\bar{r}}_k) \bm{v_k}\Big]}{\alpha_i+\beta_i+\sum_{r \in R}\Big[\; \sum_{\mathclap{\substack{\vspace{2mm}\\j:(i,j) \\ \in E_r}}}(\gamma^r_i + \gamma^{\bar{r}}_j)- \sum_{\mathclap{\substack{\vspace{2mm}\\k:(i,k) \\ \in \widetilde{E_r}}}}(\delta^r_i + \delta^{\bar{r}}_k)\Big]}
\end{flalign}
As already described earlier, optimizing the global loss may lead to large shifts of single vectors.
Moreover, as seen in \eqref{eq:convex_condition} the loss function might turn non-convex with the wrong parameter configuration.
More specifically, $\delta$ has to be much lower than $\alpha$.\\
In order to avoid those obstacles, we also derive an alternative update function which defines the values of the embedding $\bm{v_1},\ldots,\bm{v_n}$ by a series:
\begin{flalign}\label{eq:series-update}
	\bm{v_i} =& \frac{\alpha_i\bm{v_i'} + \beta_i\bm{c_i} +  \sum_{r \in R}\Big[ \sum_{\mathclap{\substack{\vspace{2mm}\\ j:(i,j) \\ \in E_r}}} \gamma^r_i \bm{v_j} - \sum_{\mathclap{\substack{\vspace{2mm}\\k:(*,k) \\ \in E_r}}} \delta^r_i \bm{v_k}\Big]}{\Big|\Big|\alpha_i\bm{v_i'} + \beta_i\bm{c_i} +  \sum_{r \in R}\Big[ \sum_{\mathclap{\substack{\vspace{2mm}\\ j:(i,j) \\ \in E_r}}} \gamma^r_i \bm{v_j} - \sum_{\mathclap{\substack{\vspace{2mm}\\k:(*,k) \\ \in E_r}}} \delta^r_i \bm{v_k}\Big]\Big|\Big|}
\end{flalign}
By the division of the length of the vector, the series is ensured to be bounded.
Thus, the parameter setting is less restricted.
Instead of using $\widetilde{E_r}$ as in \eqref{eq:iterative1}, the difference between every vector and the centroid of all target vectors in the relation $E_r$ is calculated.
This simplifies the calculation which also leads to large performance improvements as shown in Section~\ref{sec:performance} and produces good results as one can see in the ML tasks in the evaluation.
\subsection{Retrofitting Algorithm}
\label{sec:retrofitting_algorithm}
The retrofitting algorithm can be expressed as a set of matrix operations that can be solved efficiently.
We update all vectors at once using a recursive matrix equation in a similar way as this was done in the retrofitting variant of~\cite{speer2016ensemble}.
$\Psi(W)$ can be minimized by iteratively calculating $W^k$ according to \eqref{eq:min_psy}.
\begin{flalign}
	 \label{eq:min_psy}
  \nonumber
	&W_R = \sum\limits_{r \in R} \Big[ ((\gamma^r_{ij})+ (\gamma^{\bar{r}}_{ij})^T) - ((\delta^r_{ij})+(\delta^{\bar{r}}_{ij})^T) \Big] W^k \\
	\nonumber
  &W' = \alpha W^0 + \beta\bm{c} + W_R \\
	\nonumber
	&D = \mathit{diag}\Big(\alpha + \beta + \sum\limits_{r \in R}\Big[\; \sum_{\mathclap{\substack{\vspace{2mm}\\j:(i,j) \\ \in E_r}}}(\gamma^r_i + \gamma^{\bar{r}}_j)- \sum_{\mathclap{\substack{\vspace{2mm}\\k:(i,k) \\ \in \widetilde{E_r}}}}(\delta^r_i + \delta^{\bar{r}}_k)\Big]\Big) \\
  \nonumber
  &W^{k+1} = D^{-1}W' \\
	&\bm{c} = (\bm{c_1},\ldots, \bm{c_n})\;\;\;\alpha = (\alpha_1,\ldots, \alpha_n)\;\;\;\beta = (\beta_1,\ldots, \beta_n)
\end{flalign}
The matrices $(\gamma^r_{ij})$ and $(\delta^r_{ij})$ are derived from $\gamma_i$ and $\delta_i$ as defined in Section~\ref{subsec:paramaterconf}.\\
Likewise, we can use matrix operations for the alternative series based approach (see Eq.~\ref{eq:series-update}):
\begin{flalign}
  \label{eq:alternative}
	\nonumber
	&W_R = \sum\limits_{r \in R} \Big[ (\gamma^r_{ij}) - (\delta^r_{ij}) \Big] W^k \\
	\nonumber
	&W' = \alpha W^0 + \beta\bm{c} + W_R \\
	\nonumber
	&D = \mathit{diag}((W'\odot W')\mathds{1}_n) \\
	&W^{k+1} = D^{-\frac{1}{2}}W'
\end{flalign}
Usually, a low number of iterations is sufficient.
For our experiments, we set it to a fixed number of 20.
\subsection{Parameter Configuration}
\label{subsec:paramaterconf}
\begin{figure*}[t]
\begin{subfigure}[c]{0.24\textwidth}
 \center{
 \includegraphics[width=\linewidth]{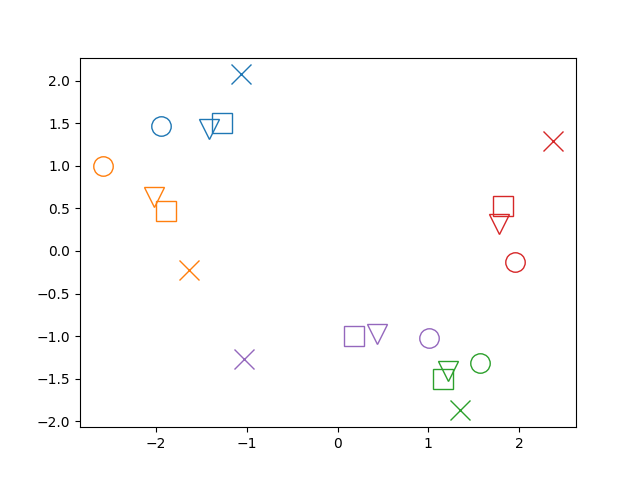}
 \tiny$\beta=1, \gamma=2, \delta=1$}
 \caption{Influence of $\alpha = 1, 2, 3$}\label{fig:alpha}
\end{subfigure}
\begin{subfigure}[c]{0.24\textwidth}
 \center{
 \includegraphics[width=\linewidth]{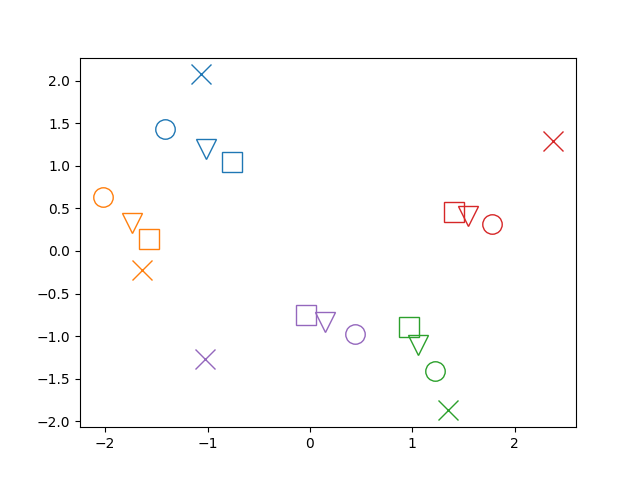}
 \tiny$\alpha = 2, \gamma=2, \delta=1$}
 \caption{Influence of $\beta= 1,2,3$}\label{fig:beta}
\end{subfigure}
\begin{subfigure}[c]{0.24\textwidth}
 \center{
 \includegraphics[width=\linewidth]{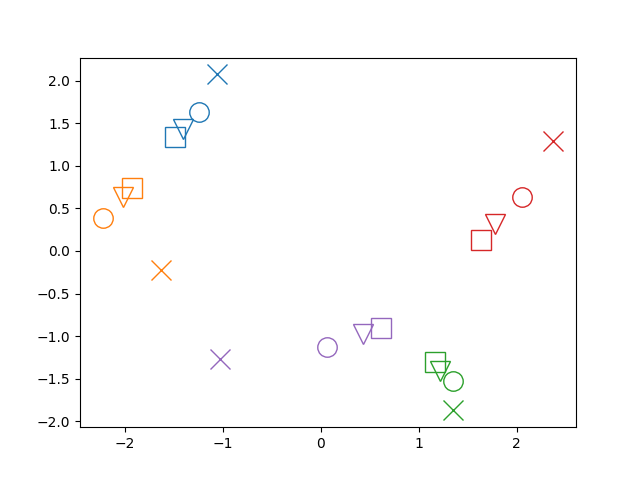}
 \tiny$\alpha = 2, \beta = 1, \delta=1$}
 \caption{Influence of $\gamma = 1,2,3$}\label{fig:gamma}
\end{subfigure}
\begin{subfigure}[c]{0.24\textwidth}
 \center{
 \includegraphics[width=\linewidth]{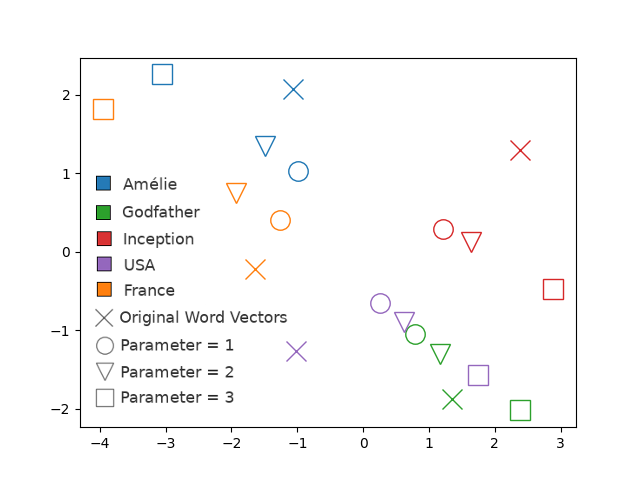}
 \tiny$\alpha = 2, \beta = 1, \gamma = 3$}
 \caption{Influence of $\delta = 0,1,2$}\label{fig:delta}
\end{subfigure}
\caption{Examples for Different Hyperparameter Settings}
\label{fig:hyper_params}
\end{figure*}
To configure the hyperparameters, we define four global parameters $\alpha$, $\beta$, $\gamma$, and  $\delta$ from which we derive the setting of all other parameters.
All $\alpha_i$ values are equivalent to the global value $\alpha$.
The other parameters are set to values that depend on the dataset.
Every embedding $\bm{v_i}$ has one categorial connections and $|R_i|$ relational connections.
Accordingly, we set the $\beta_i$ values by taking into account the number of relationship types $|R_i| + 1$ (including the categorial relationship) for weighting the influence of the category information.
Respectively, the values $\gamma_i$ are weighted by $|R_i| + 1$ and the relation-group-specific outdegree values $\mathit{od}_r(i)$:
\begin{flalign}
  \nonumber
  \beta_i = \frac{\beta}{|R_i|+1} &&\\
  \nonumber
  \gamma^r_{ij} = \begin{cases}
  \gamma_i^r = \gamma/(\mathit{od}_r(i)\cdot(|R_i|+1)) & (i,j) \in E_r \\
  0 & \mathit{otherwise}
  \end{cases} &&\\
  \mathit{od}_r(i) = |\{j:(i,j)\in E_r\}|
\end{flalign}
To fulfill condition \eqref{eq:convex_condition} we set the $\delta$ values dependent on the maximal number of relation types $\mathit{mr}$ and the maximal number of relations $\mathit{mc}$ of any node in $r$.
\begin{flalign}
	\nonumber
	\delta^r_{ij} =& \begin{cases}
  \delta_i^r = \frac{\delta}{\mathit{mc}(r) \mathit{mr}(r)} & (i,j) \in \widetilde{E_r} \\
  0 & \mathit{otherwise}
  \end{cases}&&\\
	\nonumber
	\mathit{mr}(r) =& \mathit{max}(\{|R_i|+1 | (i,j) \in E_r \cup E_{\bar{r}}\})&&\\
	\mathit{mc}(r) =& \mathit{max}(|\{i:(i,j)\in E_r\}|, |\{j:(i,j)\in E_r\}|)&&
\end{flalign}
%
%
Since we are not restricted by \eqref{eq:convex_condition} for the parameters of the alternative approach, we set the $\delta$ values as follows in this case:
\begin{flalign}\label{eq:delta_rn}
	\delta^r_{ij} = \begin{cases}
  \delta_i^r = \frac{\delta}{|\{j:(i,j)\in E_r\}|\cdot (|R_i|+ 1)} & \exists j: (i,j) \in E_r \\
  0 & \mathit{otherwise}
  \end{cases}
\end{flalign}
%
\textbf{Example.} The influence of the hyperparameters is visualized in Figure~\ref{fig:hyper_params}:
We trained 2-dimensional embeddings for a small example dataset containing three movies and the country where those movies have been produced.
Accordingly, there are two categories (movie and country) and one relation group (see Section~\ref{sec:db_rel_ext}).
``Am\'elie'' was produced in ``France'', the other movies in the ``USA''.
We set the hyperparameters $\alpha, \beta, \gamma$, and $\delta$ to different values and performed the relational retrofitting.\\
As shown in Figure~\ref{fig:alpha}, the learned embeddings stay closer to their original embeddings when the $\alpha$ values increasing.
Higher values of $\beta$ make it easier to cluster the categories from each other, e.g. reduce the distances between the movie vectors of ``Inception'' (red), ``Godfather'' (green), and ``Am\'elie'' (blue).
The $\gamma$ value controls the influence of relational connections.
This brings the representations of text values which share a relation closer together.
The $\delta$ factor causes vectors with different relations to separate and thus prevent concentrated hubs of vectors with different semantic.
One can see in Figure~\ref{fig:delta} how $\delta=0$ causes all vectors to concentrate around the origin of the coordinate system.
If $\delta$ is set to a high value like $\delta = \alpha = 2$, the algorithm places the vectors far from the origin of the coordinate system.
However, related text values still get assigned to similar representations.
In the example, the retrofitting algorithm is still converging for this configuration.
However, for higher values of $\delta$, some vectors will drift away more and more with every iteration.
\subsection{Optimization}
\label{sec:optimization}
The maximal number of relations in $E_r$ is defined by $|C_s|\cdot|C_t|$ where $|C_s|$ and $|C_t|$ refer to the cardinality of the columns involved in $r$.
However, in practice most relations are much smaller, hence $|\widetilde{E_r}| >> |E_r|$.
To optimize the extensive calculation of $((\delta^r_{ij})+(\delta^{\bar{r}}_{ij})^T)W^k$ in \eqref{eq:min_psy} we utilize the following condition:
\begin{flalign}
	\nonumber
	(((\delta^r_{ij})+(\delta^{\bar{r}}_{ij})^T)W^k)_{i,*} = 2\Big(\bm{\hat\delta^r}W^k-\hat\delta^r_i\sum_{\mathclap{j:(i,j)\in E_r}} \bm{v_j}\Big) \\
	\nonumber
\bm{\hat\delta^r} = (\hat\delta^r_1, \ldots, \hat\delta^r_n) \\
\hat\delta^r_i = \begin{cases}
		\frac{1}{\mathit{mc}(r)\cdot \mathit{mr}(r)}\; & \exists k: (i,k) \in E_r \\
		0 & \mathit{otherwise}
  \end{cases}
\end{flalign}
By calculating the vector $\bm{\hat\delta^r}W^k$ once and subtracting the centroid of the small number of related embeddings of every embedding $\bm{v_i}$, we can speed up the calculation for most cases.\\
For the alternative approach in \eqref{eq:alternative}, the following condition holds:
\begin{flalign}
	((\delta^r_{ij})W^k)_{i,*} = \sum_{\mathclap{k:(*,k)\in E_r}}\delta^r_i\bm{v_k} = \delta^r_i \sum_{\mathclap{k:(*,k)\in E_r}}\bm{v_k}
\end{flalign}
Here $\sum_{k:(*,k)\in E_r}\bm{v_k}$ can be precomputed since it is equivalent for every vector in $W^k$.
\subsection{Node Embeddings}
\label{sec:nodeembedding}
Node embeddings \cite{perozzi2014deepwalk,cai2018comprehensive,goyal2018graph} are frequently used to create vector representations for nodes in graphs to apply ML on them.
Most frequently, those embedding techniques try to capture certain properties of the neighborhood of nodes in spatial relations of the node vectors.
In this work, we use the node embedding technique DeepWalk~\cite{perozzi2014deepwalk}, that has already been successfully applied for data integration tasks~\cite{iwata2018unsupervised}.
DeepWalk generates paths of random walks on $G$ on which a Skip-Gram~\cite{NIPS2013_5021} model is trained.
Given the graph representation of the data\-base text value (see Section~\ref{sec:graphgeneration}) node embedding techniques can be directly applied without any additional effort.
For our work, we use the embeddings generated by DeepWalk in two ways: first, as a strong baseline in our evaluation (see Section~\ref{sec:competitors}).
Second, we follow the notion of \cite{goikoetxea2016single} that showed that word embeddings can be combined with node embeddings with simple vector combination methods to improve intrinsic word similarity tasks.
Specifically, it is shown that a combination of word embeddings and node embeddings captures the human notion of similarity more accurately than word embeddings itself.
During testing several combination methods, we decided to use concatenations of both embeddings since it shows good improvements across different ML tasks.
%
\section{Evaluation}
\label{sec:eval}
Word embeddings can be evaluated by intrinsic and extrinsic tasks~\cite{DBLP:journals/corr/JastrzebskiLC17}:
Intrinsic tasks evaluate the capability of embeddings to capture semantic or syntactic properties of words, such as word similarity.
In contrast, extrinsic tasks evaluate the benefit of an embedding for a certain problem such as classification and regression.
Since we focus on a vector representation ready-to-use for ML tasks, we concentrate just on extrinsic task and omit intrinsic tasks.
This decision is also supported by previous studies \cite{schnabel2015evaluation, faruqui2015sparse} showing a low correlation between intrinsic and extrinsic evaluation results.\\
We integrate \emph{RETRO} on top of PostgreSQL.
Given an initial configuration including the connection information for a database and the hy\-perparameter configuration, \emph{RETRO} fully automatically learns the retro\-fitted embeddings and adds them to the given database.
For our evaluation we use two real-world datasets, described in Section~\ref{sec:datasets}.
Given the embeddings generated by \emph{RETRO} we evaluate the accuracy in real-world classification, imputation (both Section~\ref{sec:classify}), link prediction (Section~\ref{sec:link-pred}), and regression tasks (Section~\ref{sec:regression}).
We further show that our relational embeddings are not only able to produce good results in various ML tasks but can also contend with specific state-of-the-art competitors (Section~\ref{sec:competitors}).
\subsection{Datasets}
\label{sec:datasets}
\begin{table}[t]
	{\centering
	\def\arraystretch{1.5}
	\begin{tabular}{| l | l | l |}
		 \hline
 & \textbf{TMDB} & \textbf{Google Play} \\
 \hline
  Tables & $8 (+7\text{*})$ & $6 (+1\text{*})$ \\
  Unique Text Values & $493,751$ & $27,571$ \\ \hline
\end{tabular}}\\
\small{* tables which only express n:m relations}
\caption{Dataset Properties}
\label{tbl:datasets}
\end{table}
%
We selected two real-world datasets: The Movie Database (TMDB)\footnote{\tiny{\url{https://www.kaggle.com/rounakbanik/the-movies-dataset}}} and the Google Play Store Apps dataset\footnote{\tiny{\url{https://www.kaggle.com/lava18/google-play-store-apps/}}} which are both very popular datasets on Kaggle.
Both of them are available as CSV files and are imported in a PostgreSQL database system.
Some metadata is shown in Table~\ref{tbl:datasets}.
In the following, we expose details about the content of the datasets and the creation of the databases which are processed by \emph{RETRO} to generate embeddings.\\
\textbf{TMDB:} This dataset consists of three CSV files for movies, credits, and user ratings.
The data in the CSV file consists of columns which contain numbers, text, and data in a JSON-like format.
The JSON data format is used to represent multiple values, e.g. a movie can be assigned to multiple genres.
In order to represent those values in the database, we created additional tables and foreign key relations.
From the movie table, we imported the title, the original language, spoken languages, the story overview (short text), keywords, production countries, release dates, production companies, and the genres.
We also imported some numerical data including budgets, revenues as well as popularity scores.
Those numerical values are not assigned to a learned representation, but can be used for the evaluation of regression tasks (see Section~\ref{sec:regression}).
From the cast data we import the actors and the directors in a \emph{persons} table and create respective relations to the movie table.\\
\textbf{Google Play:} This dataset contains crawled data from the Google Play Store.
It consists out of two CSV files for apps available in the store and reviews of those apps.
We removed duplicates from the app table and reduced the entities of the table to apps which have at least one non-empty review in the review file.
In the review table, we omit duplicates and apps without any text.
The text values available in the app table are the name, a list of genres, a category, the pricing type (e.g. free, paid) and the targeted age group.
We created created a table \emph{app} containing the name as well as foreign keys to tables containing category, pricing type and targeted age.
Since the movie genre table is a n:m relation, we created an additional table to represent this relationship.
The review file contains a column for the app name and a column for the review text translated to English.
The reviews are imported in the database system in a table containing its text and a foreign key referring to the app.
Most of the review texts are quite short with a median length of $81$ characters.
%
%
\subsection{Training of Embeddings}
\begin{figure}[t]
 \centering
 \includegraphics[width=0.9\linewidth]{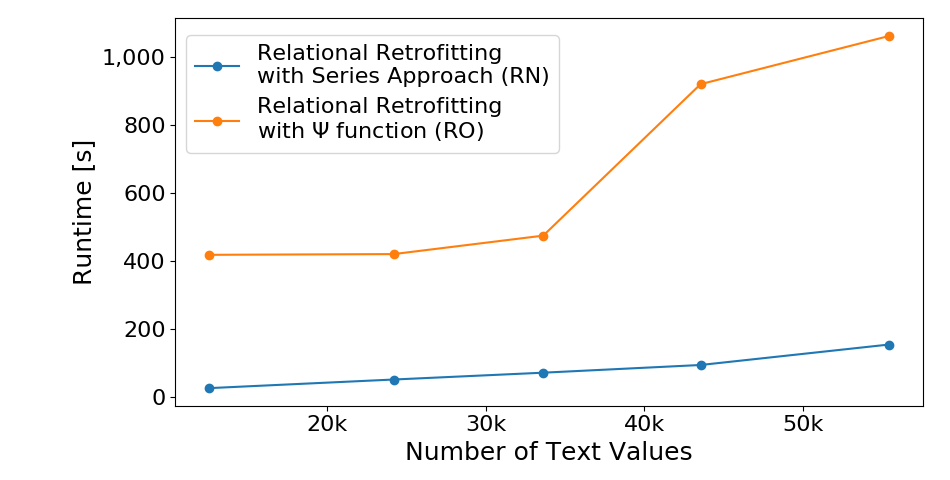}
 \caption{Runtime of Relational Retrofitting}\label{fig:perf-line-plot}
\end{figure}
\begin{figure*}[t]
\begin{subfigure}[c]{0.34	\textwidth}
 \includegraphics[width=\linewidth]{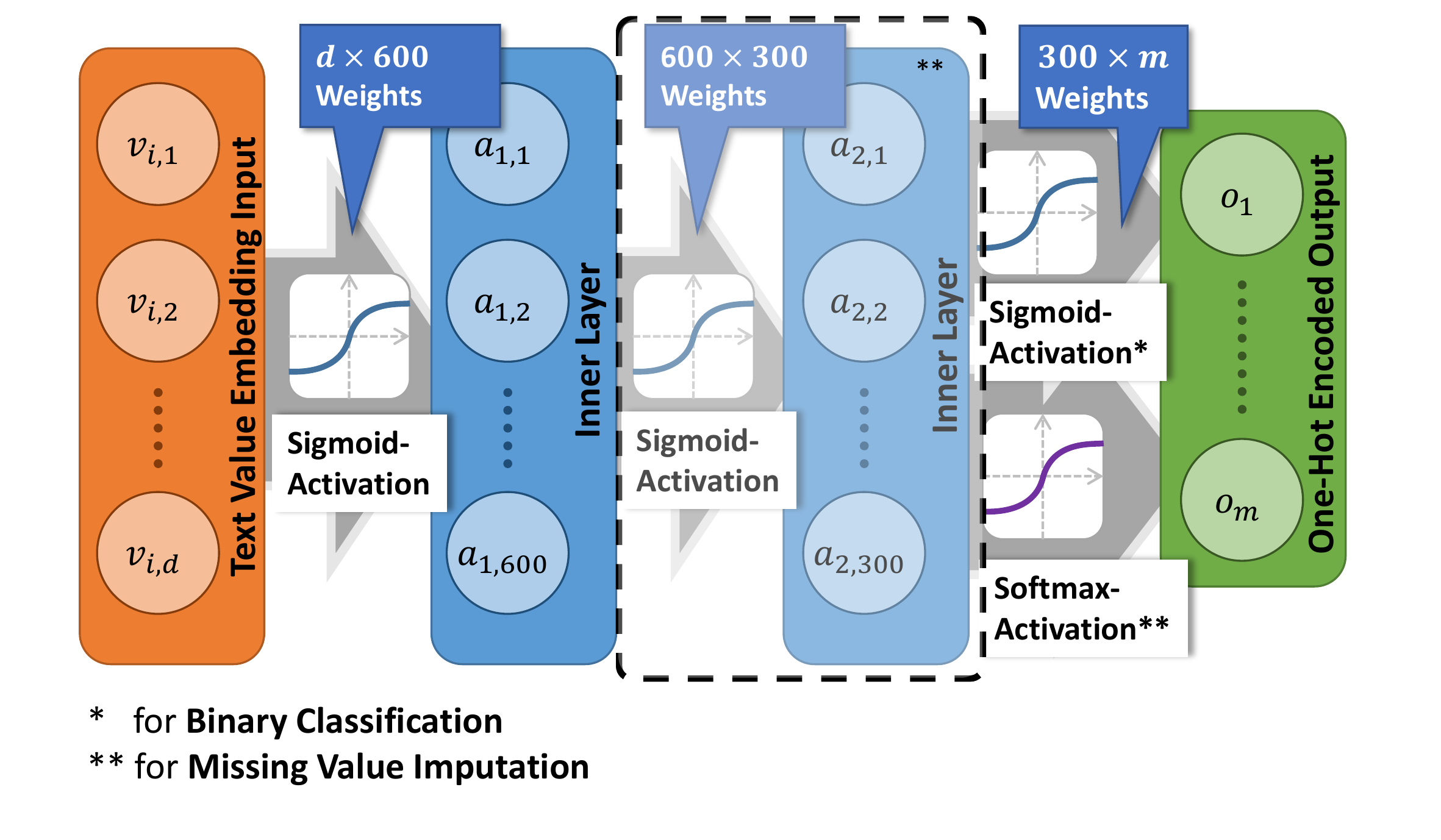}
 \caption{Classification}\label{fig:ann-classify}
\end{subfigure}
\hspace{-0.4cm}
\begin{subfigure}[c]{0.34\textwidth}
 \includegraphics[width=\linewidth]{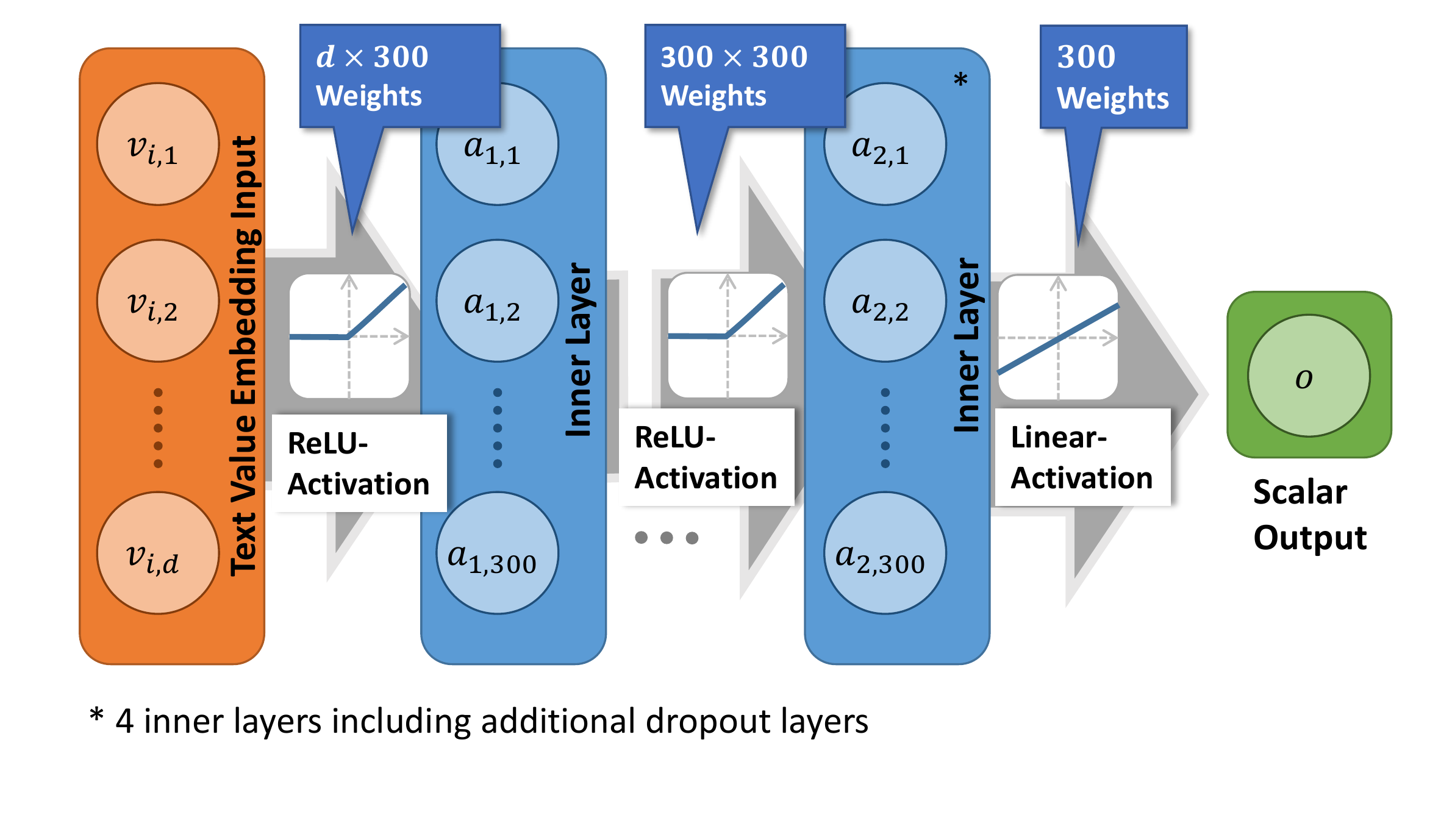}
 \caption{Regression}\label{fig:ann-regression}
\end{subfigure}
\hspace{-0.4cm}
\begin{subfigure}[c]{0.34\textwidth}
 \includegraphics[width=\linewidth]{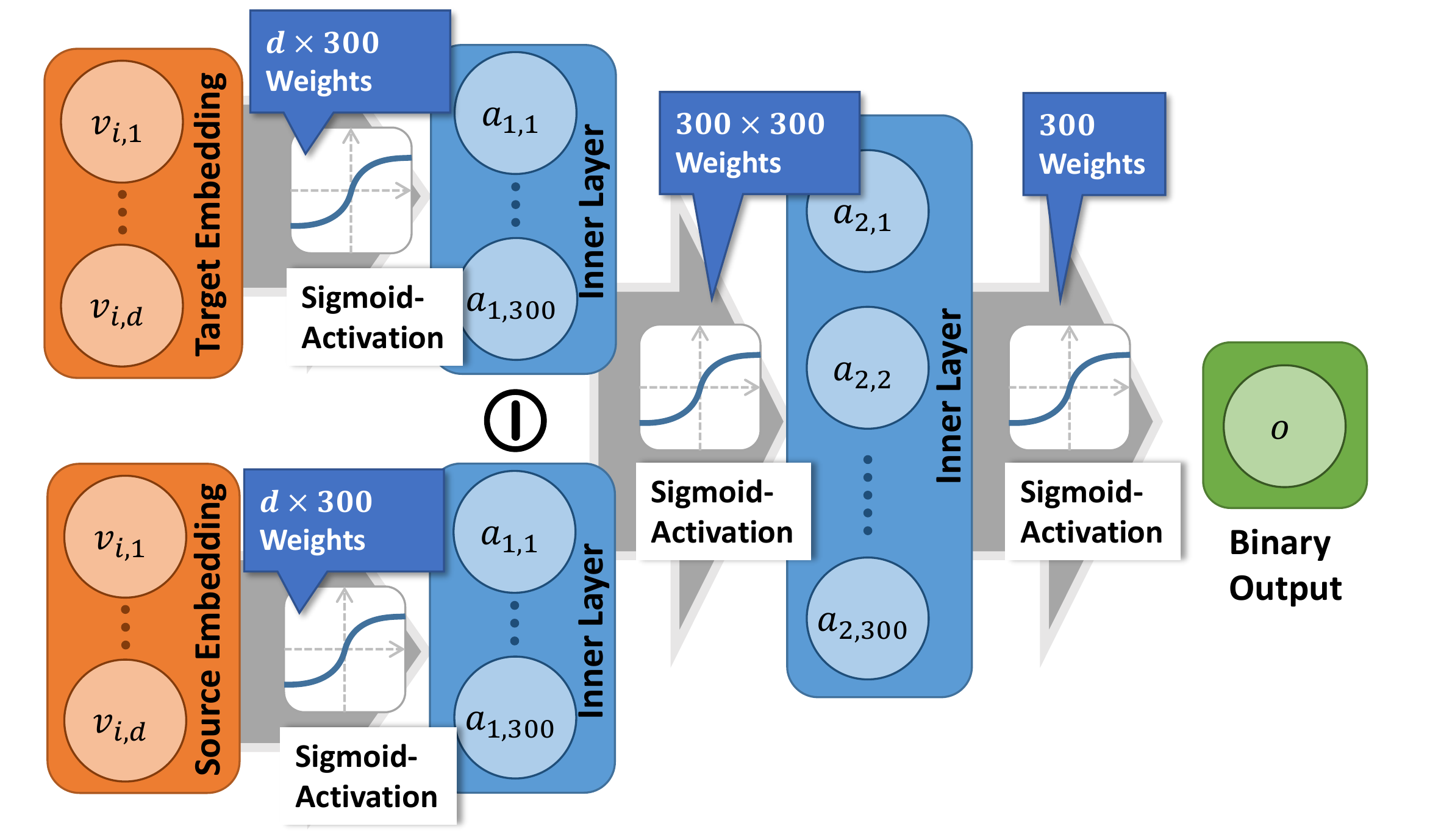}
 \caption{Link Prediction}\label{fig:ann-lp}
\end{subfigure}
\caption{ANN Architectures for the Different Extrinsic Tasks}
\end{figure*}
We trained relational embeddings for both databases with a fixed number of $10$ optimization iterations.
In order to choose a good hyperparameter configuration for the relational retrofitting approaches, we made a grid search for the binary classification and an imputation tasks.
However, for the other machine learning tasks, we choose a configuration of $\alpha = 1, \beta = 0, \gamma = 3, \delta = 3$ for the optimization function based approach (RO) and $\alpha = 1, \beta = 0, \gamma = 3, \delta = 1$ for the series based approach (RN) which turned out to be a good fit for the other tasks.
We also trained word embeddings with the basic retrofitting approach of Faruqui et al. denoted as MF, with $20$ iterations and the standard parameter configuration of $\alpha_i = 1$ and the reciprocal of the outdegree of $i$ for $\beta_i$.
We used the popular $300$ dimensional Google News embeddings\footnote{\tiny{\url{https://code.google.com/archive/p/word2vec/}}} as the base set of word embeddings for the retrofitting.
DeepWalk (DW) is trained with its standard parameters.
The dimensionality for the resulting vectors is set to $300$.
\subsection{Performance Measurements}
\label{sec:performance}
\begin{table}[t]
	\begin{tabular}{| c | c | c | c | c |}
		\hline
 & \textbf{MF} & \textbf{DW} & \textbf{RO} & \textbf{RN} \\ \hline
 \multicolumn{5}{|c|}{TMDB  } \\ \hline
 Runtime* & $7.39$ & $548.72$ & $418.13$ & $27.24$ \\
 Deviation*&+/-$0.07$ & +/-$0.83$ & +/-$1.15$ & +/-$0.51$ \\ \hline
 \multicolumn{5}{|c|}{Google Play} \\ \hline
 Runtime* & $12.23$ & $1130.63$ & $178.78$ & $35.63$ \\
 Deviation*&+/-$0.30$ & +/-$13.85$ & +/-$0.55$ & +/-$0.74$ \\ \hline
\end{tabular}
\\
\small{* values in seconds}
\caption{Runtime of Embedding Methods}
\label{tbl:runtime}
\end{table}
We execute runtime measurements for the relational retro\-fitting based on the optimization function (RO) and based on the series (RN) for different fractions of the TMDB data\-base using a single thread.
Databases of different size are generated by removing all movies with ids greater than 500, 1,000, 2,000, 4,000, and 8,000.
We removed all text values without connections to those movies, resulting in data\-bases with 12,593, 24,203, 33,628, 43,540, and 55,385 unique text values.
The results of the runtime measurements are shown in Figure~\ref{fig:perf-line-plot}.
The number of text values only express an rough estimate of the computational effort, since the execution time also depends on the number of relations and the number of relational groups.
For the TMDB dataset, the runtime of both methods increases linearly with respect to the number of text values.
Due to the simpler update function, the relational retrofitting based on the series approach (RN) is about $10$ times faster than the relational retrofitting based on the optimization function (RO) for the TMDB dataset.
To compare the training time of the different retro\-fitting approaches and DeepWalk, we measure the execution time of the different methods executed in a single thread for both datasets.
The measurements are repeated $10$ times.
Due to the very high training times of DeepWalk, we used the subset of the TMDB database with 12,593 unique text values.
The results of the measurements are shown in Table~\ref{tbl:runtime}.
As one can see, the retrofitting algorithms (MF, RO, and RN) are faster than the DeepWalk (DW) node embedding method.
As one would expect the basic retrofitting approach (MF) is even faster.
This improvement derives from the simplified modeling of database relations, leading however to lower accuracies in ML tasks as seen in the remainder of the evaluation.
Comparing the runtimes between the Google Play and the TMDB dataset, we see an increase for all methods except RO relational retrofitting.
This behavior is explained by the larger amount of relational groups in the TMDB subset compared to the Google Play dataset leading to a larger amount of comprehensive matrix multiplications which are the most time-consuming routine in the RO retrofitting.
%
%
\subsection{Competitors}
\label{sec:competitors}
\begin{figure*}[t]
\begin{subfigure}[c]{0.52\textwidth}
 \centering
 \includegraphics[width=\linewidth]{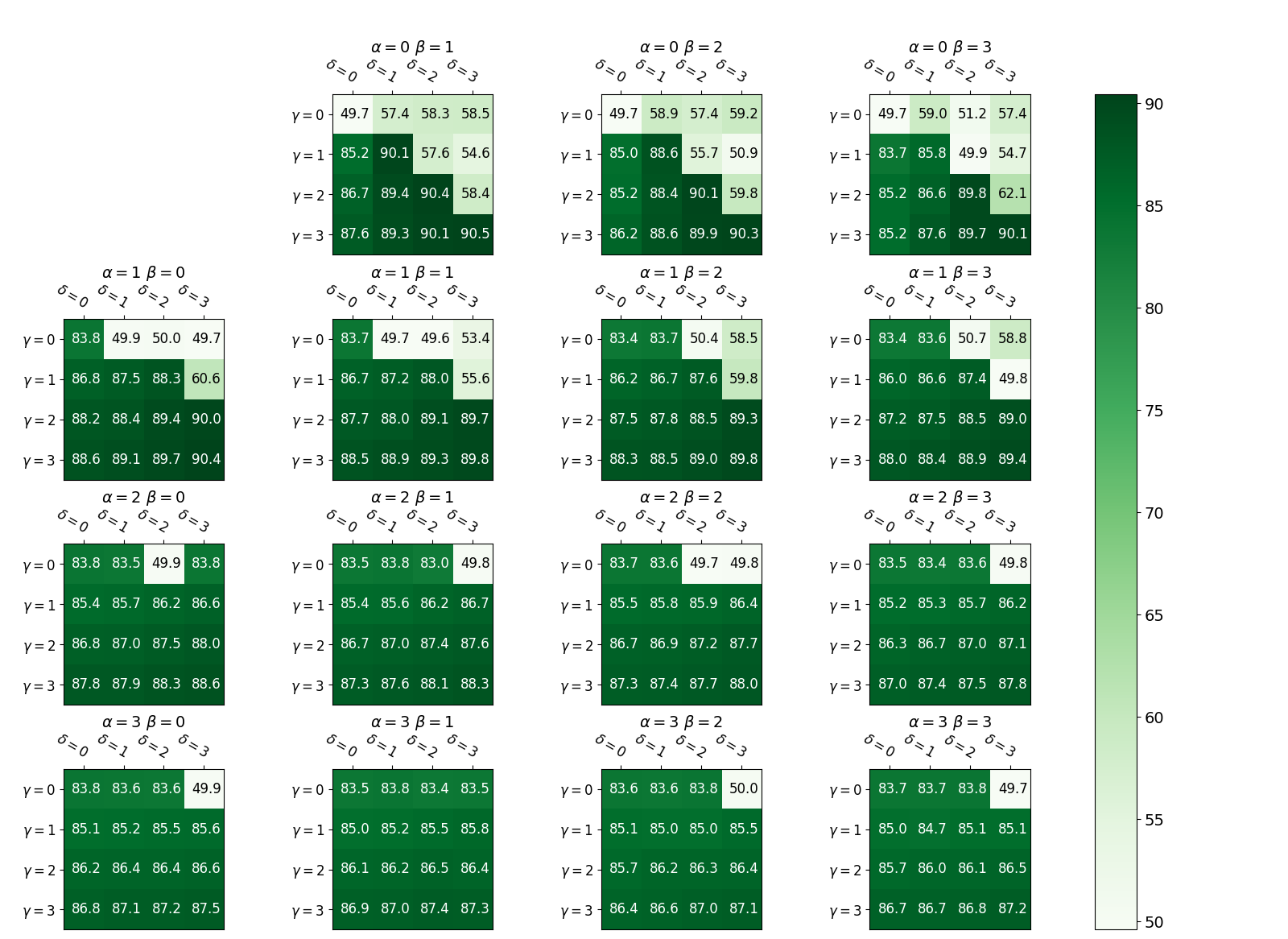}
 \caption{Only Retrofitted Vectors}\label{fig:heatmap_dc_single_ro}
\end{subfigure}
\hspace{-13mm}
\begin{subfigure}[c]{0.52\textwidth}
 \centering
 \includegraphics[width=\linewidth]{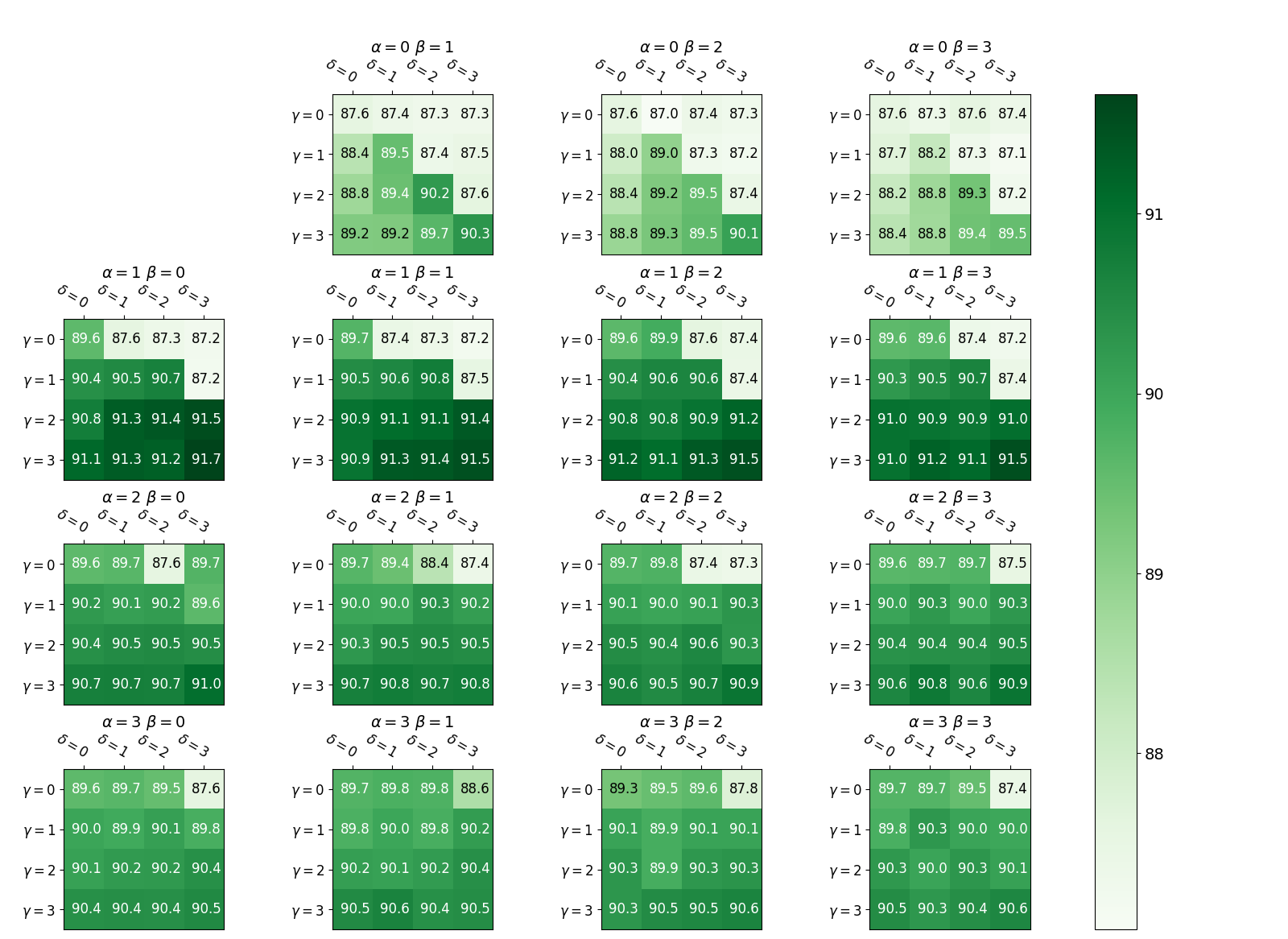}
 \hspace{-10mm}
 \caption{Retrofitted Vectors combined with DeepWalk Embeddings}\label{fig:heatmap_dc_combined_ro}
 \end{subfigure}
 \caption{Influence of Hyperparameters on Binary Classification for Relational Retrofitting with $\Psi$ Function}\label{fig:heatmaps_dc_ro}
\end{figure*}
\begin{figure*}[t]
\begin{subfigure}[c]{0.52\textwidth}
 \centering
 \includegraphics[width=\linewidth]{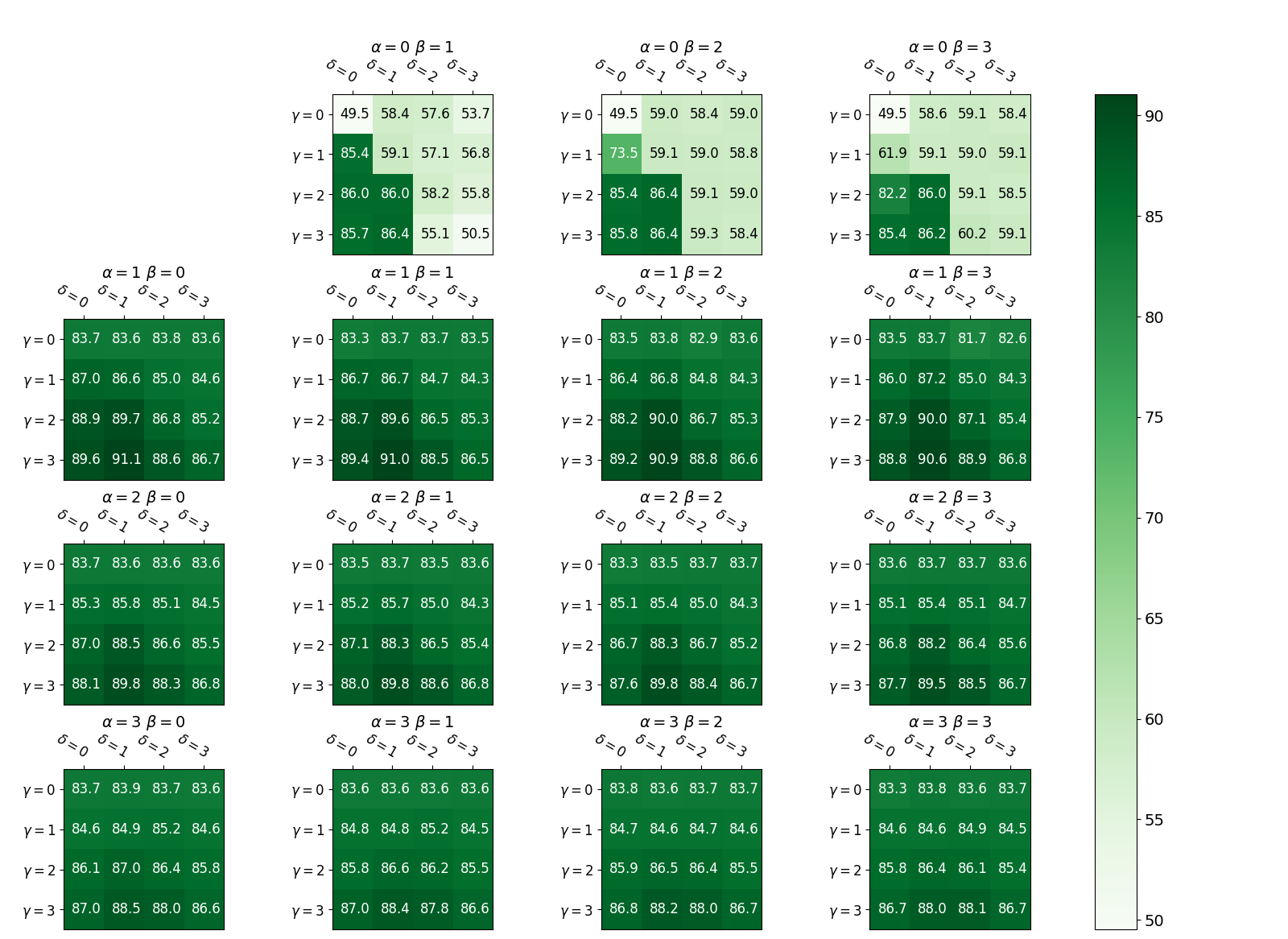}
 \caption{Only Retrofitted Vectors}\label{fig:heatmap_dc_single_rn}
\end{subfigure}
\hspace{-13mm}
\begin{subfigure}[c]{0.52\textwidth}
 \centering
 \includegraphics[width=\linewidth]{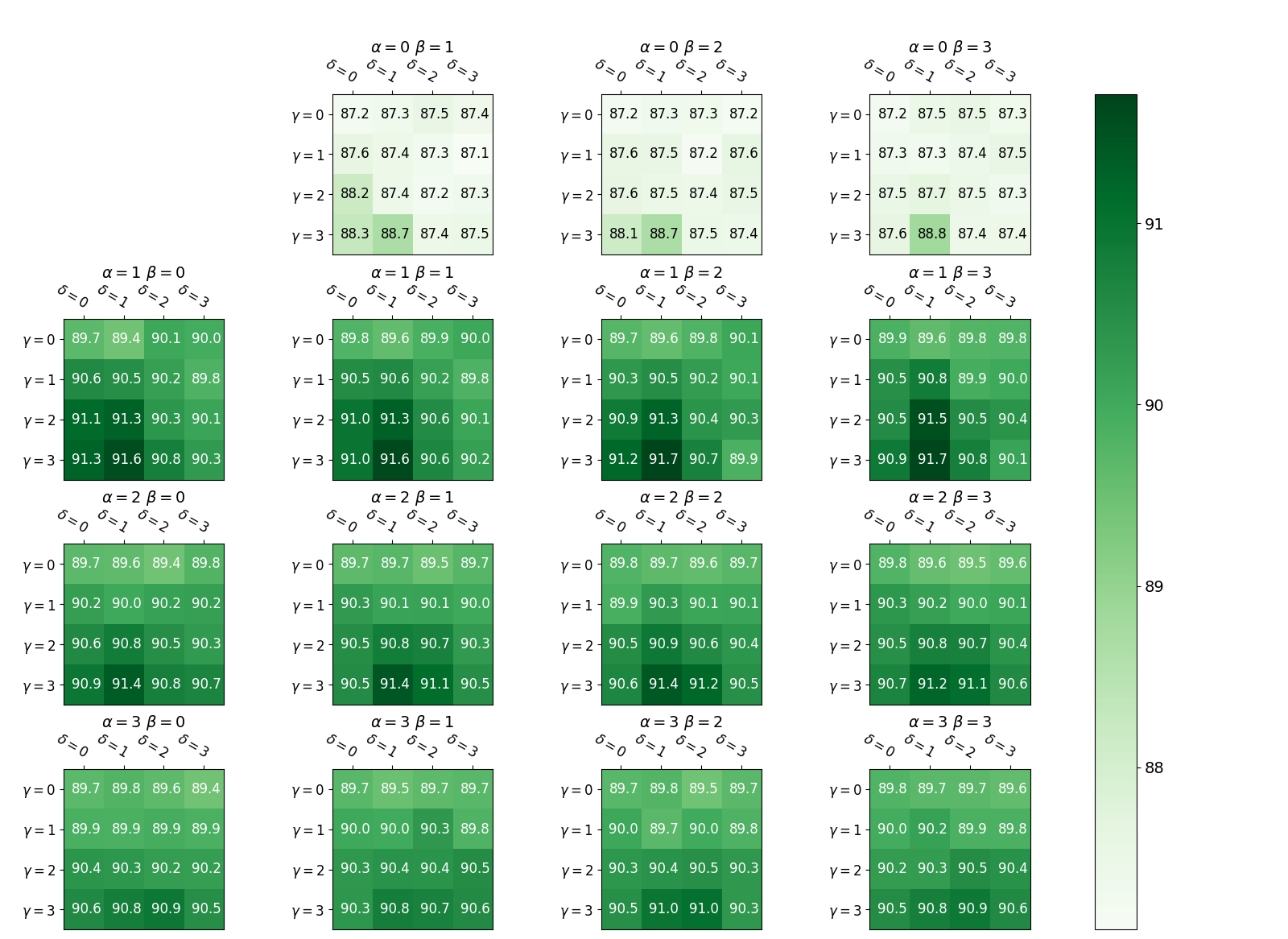}
 \hspace{-10mm}
 \caption{Retrofitted Vectors combined with DeepWalk Embeddings}\label{fig:heatmap_dc_combined_rn}
 \end{subfigure}
 \caption{Influence of Hyperparameters on Binary Classification for Relational Retrofitting with Series Approach}\label{fig:heatmaps_dc_rn}
\end{figure*}
To demonstrate the applicability of \emph{RETRO} and it's learned representations for different tasks we use following baseline approaches.\\
\textbf{DataWig:} Missing value imputation on spreadsheets with text values can be accomplished with DataWig~\cite{biessmann2018deep}.
This recently published category imputer is based on n-gram representations of text values which are utilized by LSTM neural networks to assign text values in a spreadsheet to categories.
As an input, the imputer gets a spreadsheet, a list of columns used for the imputation, a column which contains the values which should be assigned to categories, and the column which usually holds the output category.
The imputation is then trained on a sample set of rows.
We compare DataWig against imputation using our embeddings in Section~\ref{sec:classify}.\\
\textbf{Mode Imputation:} A very simple method for im\-pu\-ta\-tion is to replace a null value by the mode value (most frequently value) occurring in the column.
According to~\cite{biessmann2018deep}, most data-wrangling frameworks implement only such simple imputation methods for category imputation, especially if there is only non-numerical data given.
Mode is a very popular and often used imputation method and, thus, it also serves as a good baseline.\\
\textbf{DeepWalk:} The node embeddings generated by DeepWalk (see Section~\ref{sec:nodeembedding}) are often used for link prediction tasks and thus can be used to predict missing foreign key relations in database systems.
The prediction is usually performed using feed-forward neural networks similarly as one could use the text value embeddings proposed in this paper.
The details are described in Section~\ref{sec:link-pred}.
While we use embeddings generated with DeepWalk (DW) in combination with our embedding representations (RO+DW and RN+DW)  we also compare the performance of our learned representations (RO and RN) with the node embeddings and combination of both embedding types.
\begin{figure}[t]
 \centering
 \includegraphics[width=\linewidth]{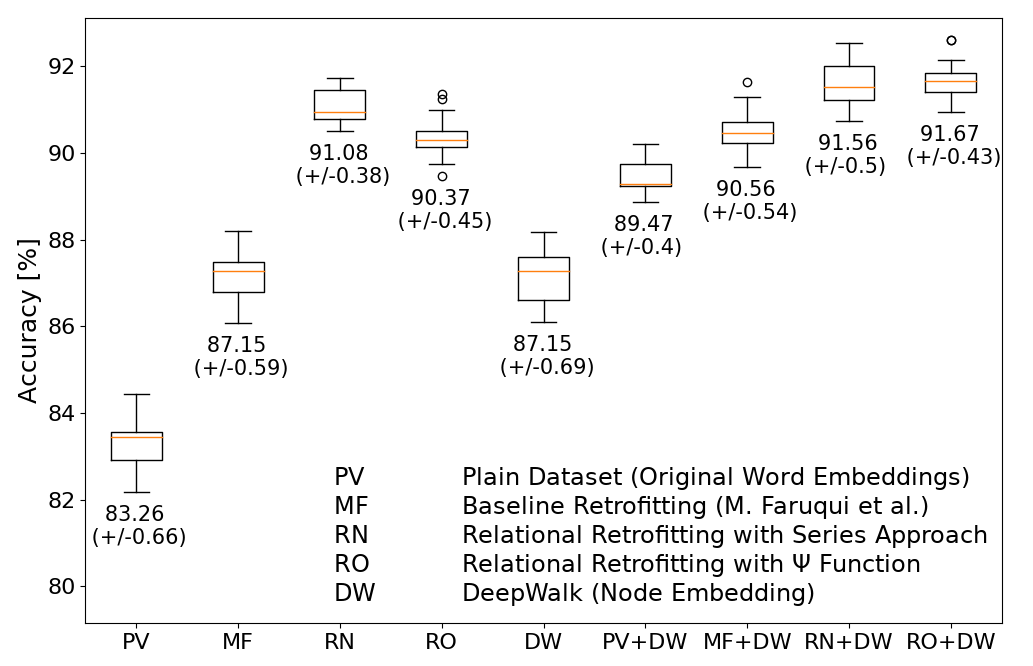}
 \caption{Binary Classification of US-American Directors with Different Embedding Types}\label{fig:classify-box-plots}
\end{figure}
\subsection{Classification}
\label{sec:classify}
As the most popular extrinsic evaluation tasks we want to perform classification.
%
In the following, we distinguish \emph{binary classification} and \emph{category imputation} (mutually exclusive classes).
For the binary classification, a classifier has to determine if a text value is assigned to a label or not.
In case of a category imputation problem, the classifier can select a label from a set of labels.
One could also solve multi-class classification problems where each text value can be assigned to multiple classes.
However, for our evaluation, we focus on the other two settings.\\
\textbf{Network Architecture:} For both task we use a feed-forward neural network (Figure~\ref{fig:ann-classify}).
The first layer has 600 and the second 300 neurons.
We used sigmoid activation functions for the fully connected inner layers.
For the binary classification, one hidden layer is sufficient.
The ANN for category imputation uses two hidden layers.
Both ANNs use a sigmoid function for the output.
In the case of the category imputation, softmax is used for the output layer.
As a loss function, we use a categorical cross-entropy loss for the category imputation and binary cross-entropy for the other cases.\\
To prevent overfitting, we added dropouts~\cite{srivastava2014dropout} and L2 regularization in case of the binary classification.
The networks are trained with the Nadam optimizer~\cite{dozat2016incorporating}.
Following common practices to prevent irregular updates of the weights, we normalize the embedding vectors before they are processed by the network.
By training the network we use $10\%$ from the training data as validation set providing the possibility to keep track of the training process.
During training, we check if the loss according to the validation set does not improve for more than 50 epochs.
If this is the case, we stop the training and select the model with the lowest loss according to the validation set and determine the accuracy for the test set.\\
\begin{figure}[t]
\vspace{-4mm}
\begin{subfigure}[c]{0.24\textwidth}
 \includegraphics[width=\linewidth]{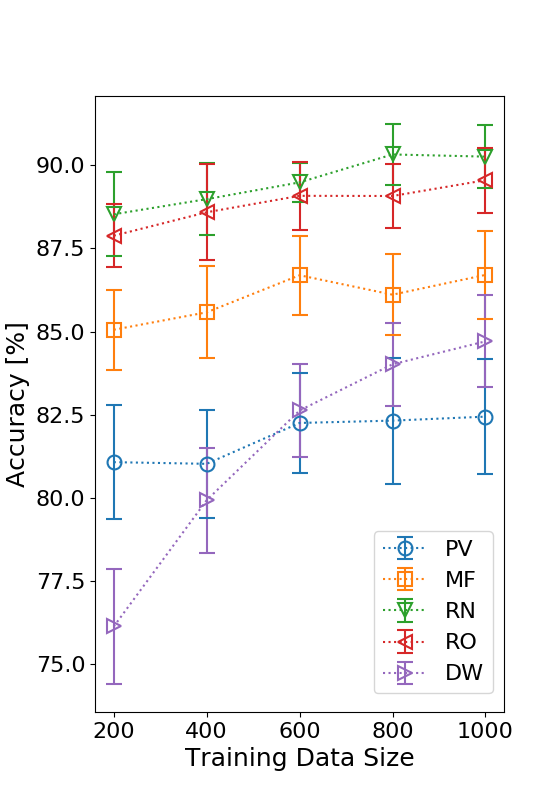}
\end{subfigure}
\hspace{-0.4cm}
\begin{subfigure}[c]{0.24\textwidth}
 \includegraphics[width=\linewidth]{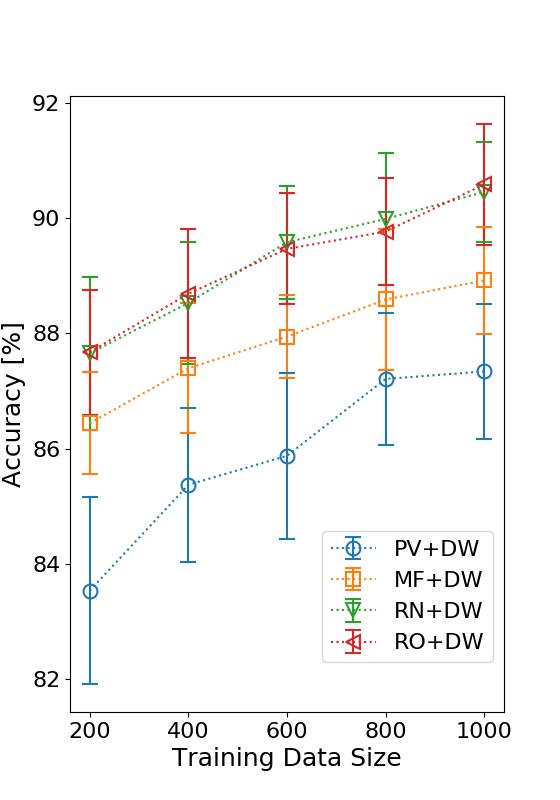}
\end{subfigure}
\caption{Classification of Birth Places of US-American Directors with Increasing Sample Size}\label{fig:classify-dc-line}
\end{figure}
\subsubsection{Binary Classification}
We use the network architecture described above to solve the following binary classification problem.
From a set of directors of the TMDB dataset, a classifier should label which directors as US-American or non-US-American.
Since this information is not available from the TMDB dataset, we extract the citizenship from Wikidata~\cite{vrandevcic2014wikidata} by using the SPARQL query service.
Thereby, we derive $33,647$ directors holding in total $37,203$ citizenships.
We omit $387$ directors which hold the US-American and another citizenship according to Wikidata because this could be considered as ambiguous.
From the remaining set of directors, the TMDB database contains $9,054$ which are considered for the classification.\\
For every embedding type, we sample $10$ times embeddings of $3,000$ US-American directors and $3,000$ non-US-American ones as a sample set.
One half of a sample set is used for training the ANN.
The other half is used as a test set to determine the accuracy.
In addition the classification is performed on embeddings derived by concatenation of DeepWalk embeddings and the other embedding types.
The retrofitting approaches are performed on the $300$ dimensional Google News word embeddings.
In order to derive a good hyperparameter configuration for the relational retrofitting, we performed a grid search.
The average accuracy values for all tested hyperparameters are shown in Figure~\ref{fig:heatmaps_dc_ro} and Figure~\ref{fig:heatmaps_dc_rn}.
One can see that high values for $\gamma$ and $\delta$ deliver good results for the optimization based approach (RO).
This suggests that relational information is more important for a good classification.
In the series based approach optimal configuration has higher values for $\gamma$ than for $\delta$, however, the influence of $\delta$ is higher than in the optimization based approach (see Equation~\eqref{eq:delta_rn}).
Configurations with a high value of $\delta$ but low values for $\alpha$ and $\gamma$ leading to non-converging configurations and worse classification results.
In combination with node embeddings, the optimal configuration has higher values for $\alpha$ and $\beta$, since the relational connections are already represented by the node embeddings.
The $\beta$ parameter has a low influence on this task since all inputs are part of the same column.
The distributions of the recognition accuracy values achieved by the classifier for RO ($\alpha = 1, \beta = 0, \gamma = 3, \delta = 3$), RN ($\alpha = 1, \beta = 0, \gamma = 3, \delta = 1$) and the other embedding types are shown in Figure~\ref{fig:classify-box-plots}.
As one can see, the best accuracy values are achieved with the relational retrofitting approaches (RN and RO), where the series approach (RN) performs slightly better than the relational retrofitting with optimization (RO).
The node embeddings (DW) are outperformed by all other types, besidse the plain word embeddings (PV) and the baseline retrofitting approach (MF) which achieves similar performance.
However, combining node embeddings with all other approaches results in an increase of their achieved accuracies above $90\%$ in all cases except the plain word vectors.\\
\begin{figure*}[t]
\begin{subfigure}[c]{0.52\textwidth}
 \centering
 \includegraphics[width=\linewidth]{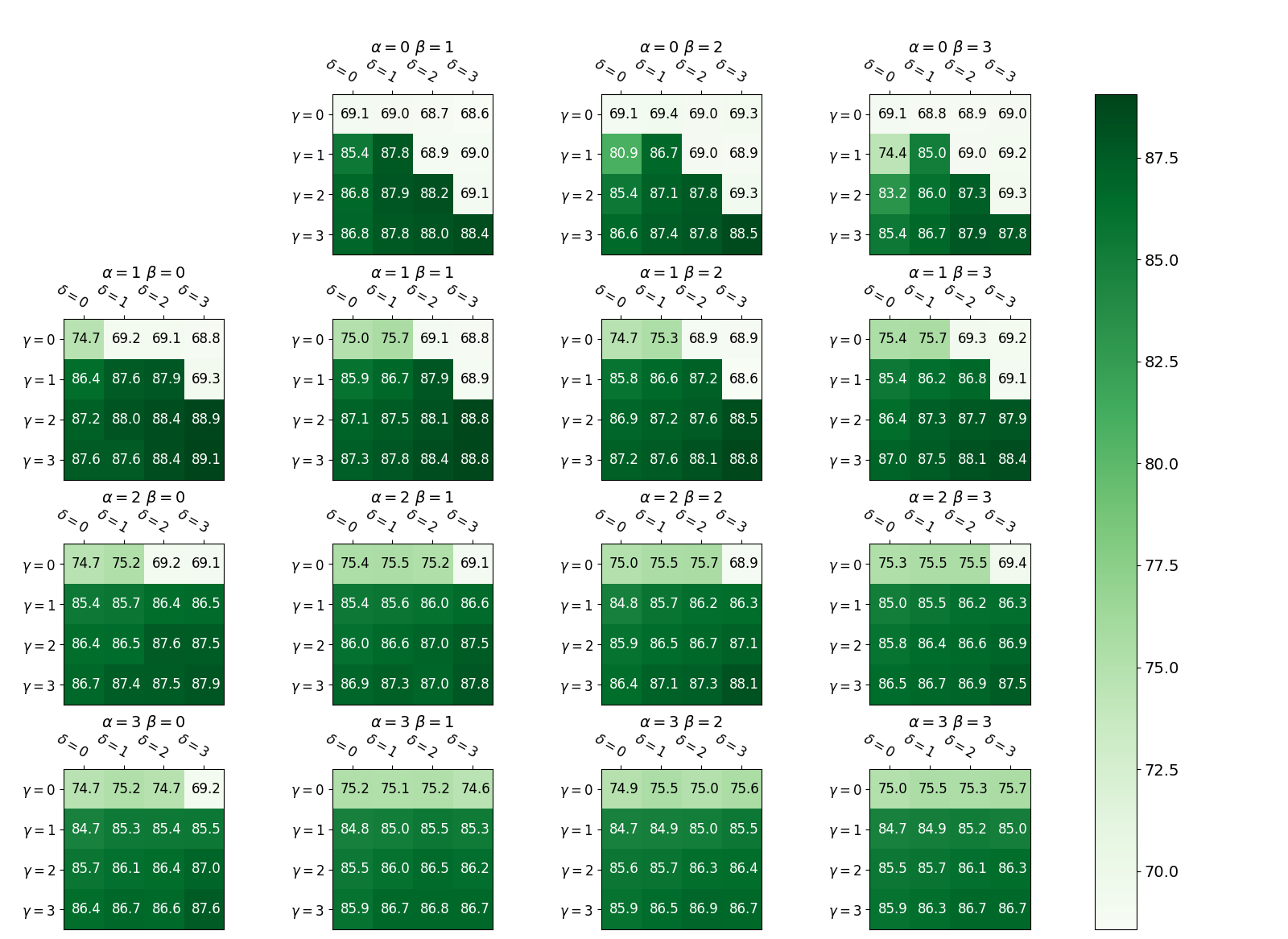}
 \caption{Only Retrofitted Vectors}\label{fig:heatmap_ol_single_ro}
\end{subfigure}
\hspace{-13mm}
\begin{subfigure}[c]{0.52\textwidth}
 \centering
 \includegraphics[width=\linewidth]{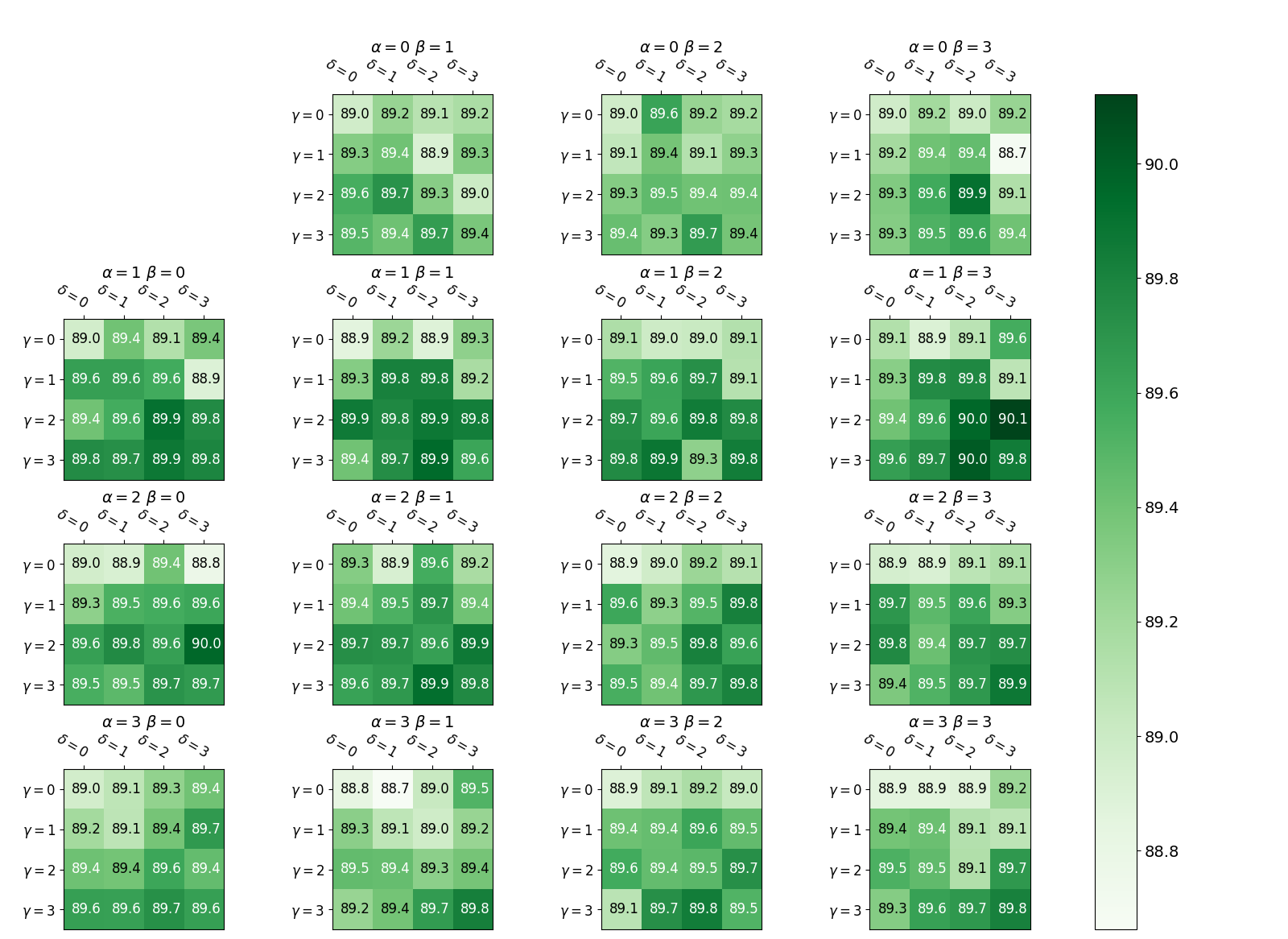}
 \hspace{-10mm}
 \caption{Retrofitted Vectors combined with DeepWalk Embeddings}\label{fig:heatmap_ol_combined_ro}
 \end{subfigure}
 \caption{Influence of Hyperparameters on Original Language Classification for $\Psi$ Function Approach}\label{fig:heatmaps_ol_ro}
\end{figure*}
\begin{figure*}[t]
\begin{subfigure}[c]{0.52\textwidth}
 \centering
 \includegraphics[width=\linewidth]{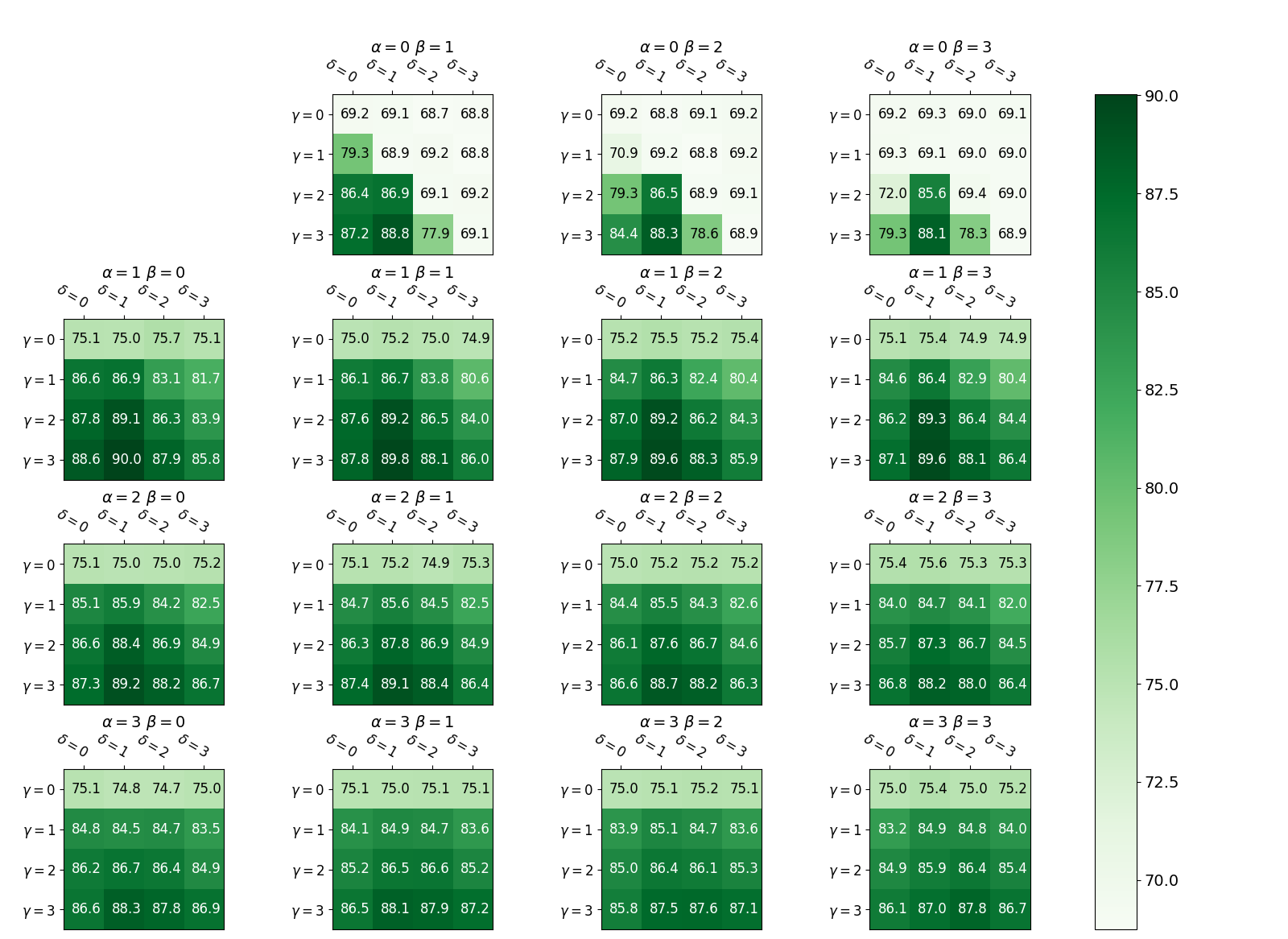}
 \caption{Only Retrofitted Vectors}\label{fig:heatmap_ol_single_rn}
\end{subfigure}
\hspace{-13mm}
\begin{subfigure}[c]{0.52\textwidth}
 \centering
 \includegraphics[width=\linewidth]{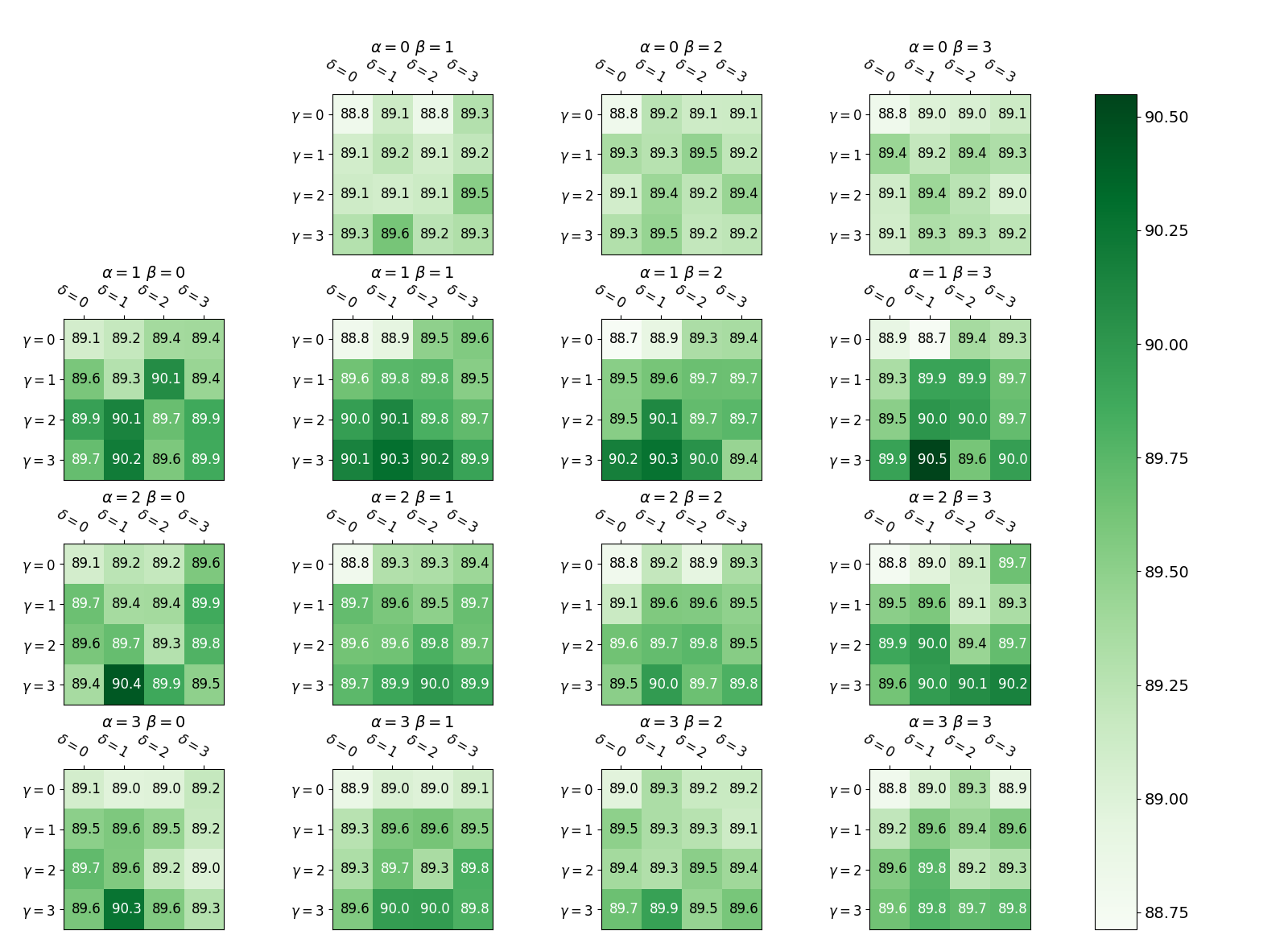}
 \hspace{-10mm}
 \caption{Retrofitted Vectors combined with DeepWalk Embeddings}\label{fig:heatmap_ol_combined_rn}
 \end{subfigure}
 \caption{Influence of Hyperparameters on Original Language Classification for Series Approach}\label{fig:heatmaps_ol_rn}
\end{figure*}
\begin{figure*}[t]
\begin{subfigure}[c]{0.48\textwidth}
 \centering
 \includegraphics[width=\linewidth]{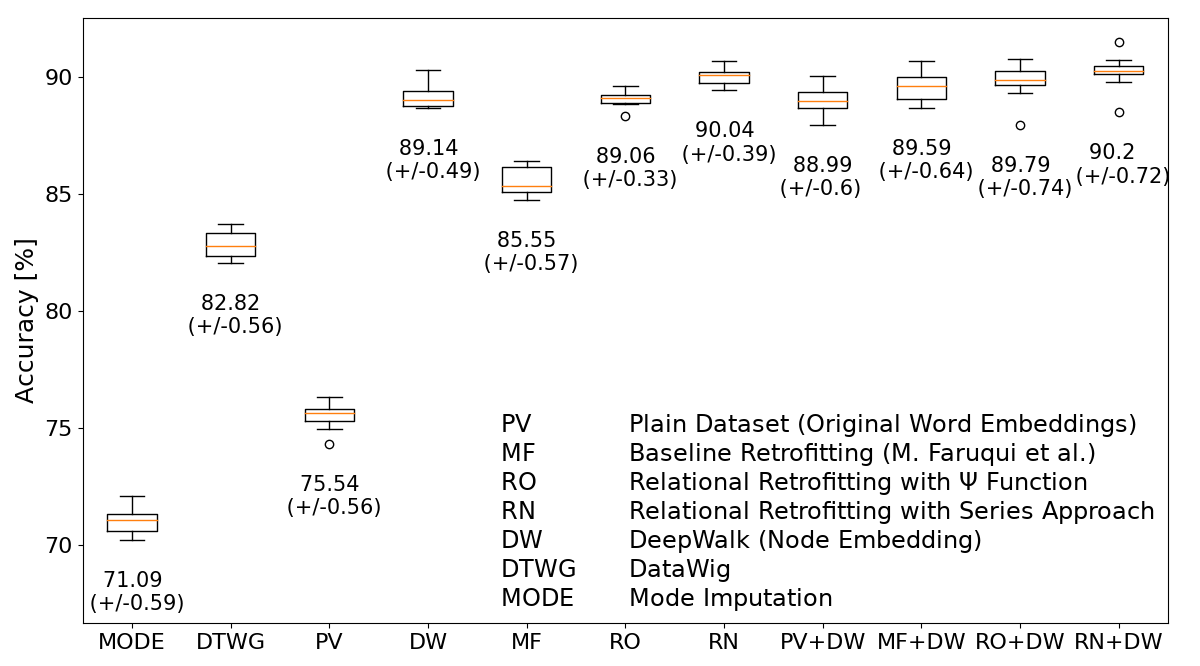}
 \caption{Imputation of Original Language Property}\label{fig:original-language}
\end{subfigure}
\hspace{5mm}
\begin{subfigure}[c]{0.48\textwidth}
 \centering
 \includegraphics[width=\linewidth]{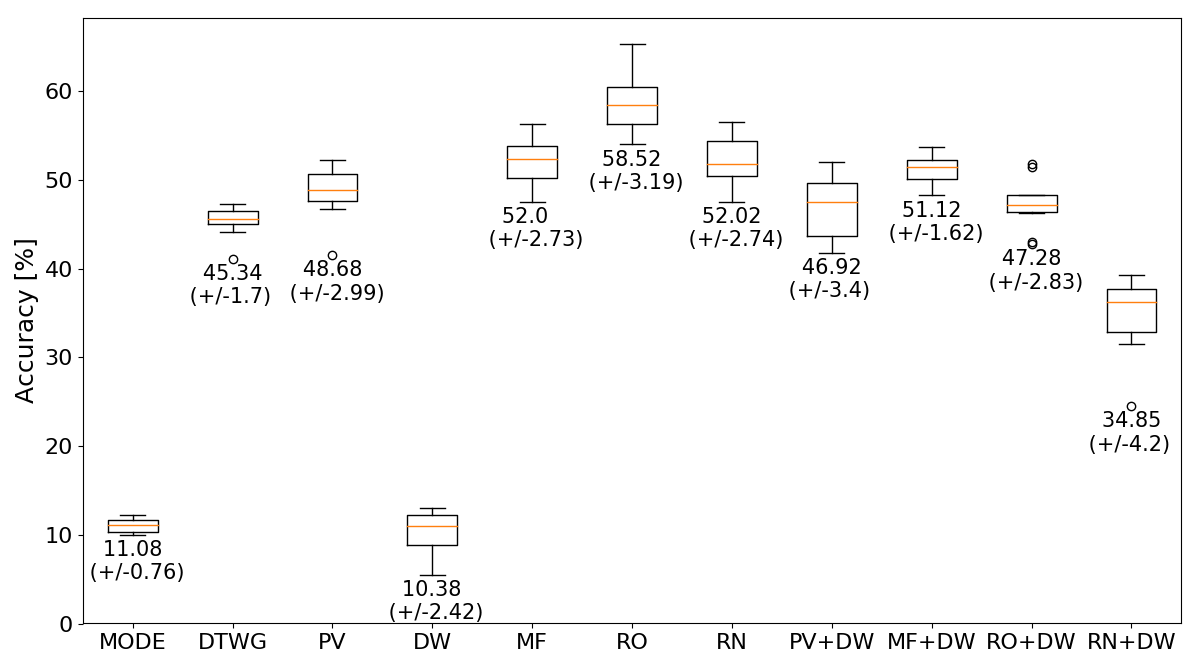}
 \caption{Imputation of App Categories}\label{fig:app-categories}
 \end{subfigure}
 \caption{Comparison of Imputation Methods}
\end{figure*}
Additionally, we ran the experiments for varying numbers of training samples.
We trained the neural network with $200$ to $1,000$ samples and validate the accuracy with $1,000$ test samples.
Since the deviation is much higher for such small sample sets, we trained the ANNs $20$ times.
The result is shown in Figure~\ref{fig:classify-dc-line}.
The influence of the training sample size is at lowest for the plain word embeddings.
If the sample sizes are small DeepWalk is getting outperformed by the plain word embeddings.
Hence DeepWalk needs a larger amount of training data to achieve comparable results.
\subsubsection{Missing Value Imputation}
As a basic data integration task, we perform missing value imputation for categories on both datasets.\\
\textbf{Imputation of Movies Language:}
In the movie dataset ``original language'' is a property of movies having exactly one value for each movie.
Subsequently, the prediction of those property values can be considered as a typical value imputation problem.
To perform the classification, we train embeddings by ignoring the ``original language'' column in the movie table.
Afterward, we train a neural network for value imputation as described above.
We use $5,000$ training samples which are split in 90\% training set and 10\% validation set and a set of $5,000$ samples to evaluate the accuracy.
Sampling, training, and evaluation are repeated 10 times for each embedding type.
The influence of the hyperparameters on the accuracy of the relational retrofitting is shown in Figure~\ref{fig:heatmaps_ol_ro} and Figure~\ref{fig:heatmaps_ol_rn}.
Here, configurations with $\alpha=1$ deliver the highest accuracy values.
The influence of $\gamma$ and $\delta$ is similar as in the binary classification task.
In addition to the embedding approaches, we apply mode imputation and DataWig with an equivalent sampling strategy.
DataWig is provided with the textual movie information in the form of a spreadsheet.
It contains all information imported into the database, except directors and actors which reside in other tables.
The accuracy values for all methods and embedding types are compared in Figure~\ref{fig:original-language}.
Since most of the movies are in the English language the mode imputation (MODE) performs quite well achieving an accuracy of $71.09\%$ in the average case.
Using plain word embeddings (PV), the accuracy is only slightly better.
The relational retrofitting approaches  RO ($\alpha = 1, \beta = 0, \gamma = 3, \delta = 3$) and RN ($\alpha = 1, \beta = 0, \gamma = 3, \delta = 1$) reach the highest accuracy values and outperform DataWig (DTWG) which only considers relations within a single spreadsheet.
DeepWalk performs similarly good as our learned representation, however, achieves even better results in combination with retrofitted embeddings.\\
\textbf{Imputation of App Categories:}
In the \emph{Google Play data\-set}, we classify the category to which an app belongs to.
For the training of the embeddings, we omit the category information and the genre relation since genre and category are often equivalent.
The apps in the dataset are grouped into $33$ categories.
Again, we compared our imputation to DataWig and the mode imputation.
Since DataWig can only be executed on singular tables we omit the review data.
We sampled $10$ times two disjunct sets of $400$ apps as training and testing sets.
The resulting accuracy values of the imputation are shown in Figure~\ref{fig:app-categories}.
Here, the mode imputation does only provide very poor accuracy values since the apps are more distributed over the category values.
DataWig achieves a clearly better performance which is similar to the performance of the plain word embeddings of the app name.
This is probably the case because DataWig might also rely strongly on the app name since all other attributes are quite unrelated to the category information.
The retrofitting based approaches RO ($\alpha = 1, \beta = 0, \gamma = 3, \delta = 3$) and RN ($\alpha = 1, \beta = 0, \gamma = 3, \delta = 1$) can utilize the reviews and achieve an up to $13\%$ more accurate result.
DeepWalk does not improve the classification in comparison to the mode imputation since classification can not be done based on relational information.
Accordingly, the embedding approaches do not improve from a concatenation with DeepWalk embeddings.
\begin{figure}[t]
 \centering
 \includegraphics[width=\linewidth]{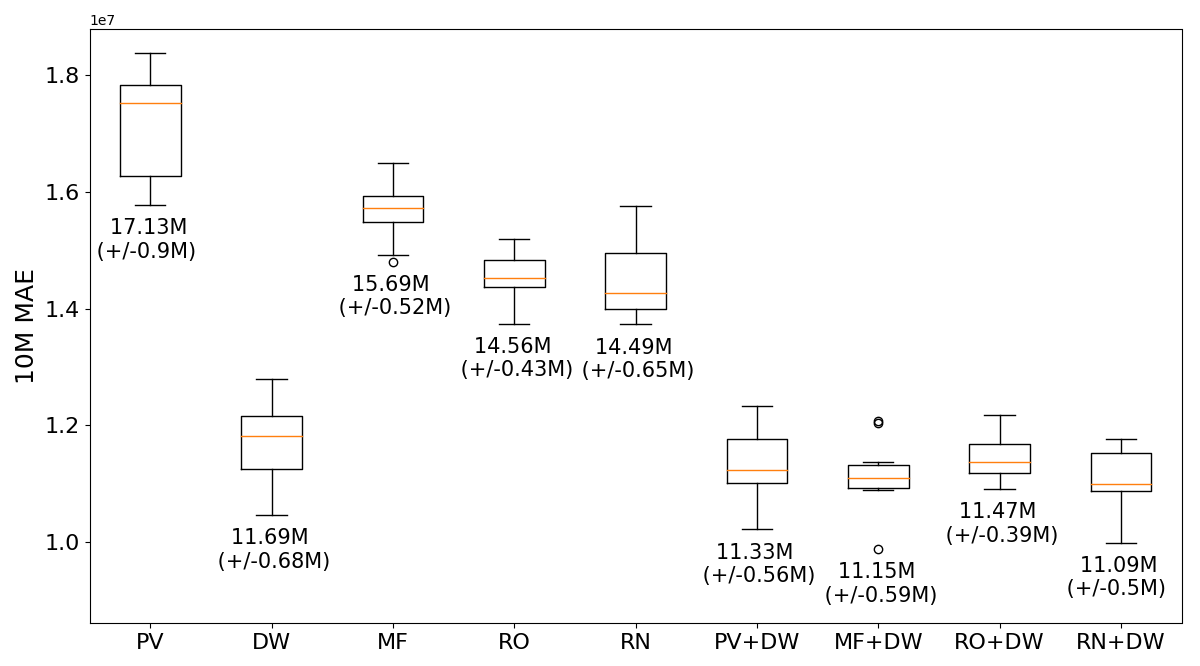}
 \caption{Regression of Budget}\label{fig:regression-budget}
\end{figure}
\subsection{Regression}
\label{sec:regression}
Regression tasks assign input data points to a related numerical value.
This could be the budget or revenue in US dollar for a movie title.
Regression tasks can also be performed with neural networks.\\
\textbf{Network Architecture:}
For the regression, an architecture is selected which is slightly different from the architecture used for binary classification problems.
Instead of sigmoid activation functions, we use ReLU functions and a linear output function since result values could be outside of the interval $[0,1]$.
Moreover, we choose a deeper network architecture with 4 inner layers and therefore higher representational capabilities which delivers during testing higher accuracy than shallow networks.
However, the simple ReLU activation functions allow fast training of the network.
The loss used for the training is the \emph{Mean Absolute Error (MAE)} function.\\
\textbf{Regression of Budget:}
The regression is executed to predict the budget for the production of movie titles.
We sampled a training dataset of $9,000$ movie titles and $1,000$ movies for testing.
We ran those experiments $10$ times.
Figure~\ref{fig:regression-budget} shows the MAE for the regression.
One can see that the node embeddings perform significantly better than the text-based embeddings.
This suggests, that relational information is prevalent over semantic features of the text values.
Relational retrofitting performs slightly better than the baseline approach.
In combination with the node embeddings, the error is similarly high for all text-based embedding but slightly lower than the error when only node embeddings are used.
\subsection{Link Prediction}
\label{sec:link-pred}
\begin{figure}[t]
 \centering
 \includegraphics[width=\linewidth]{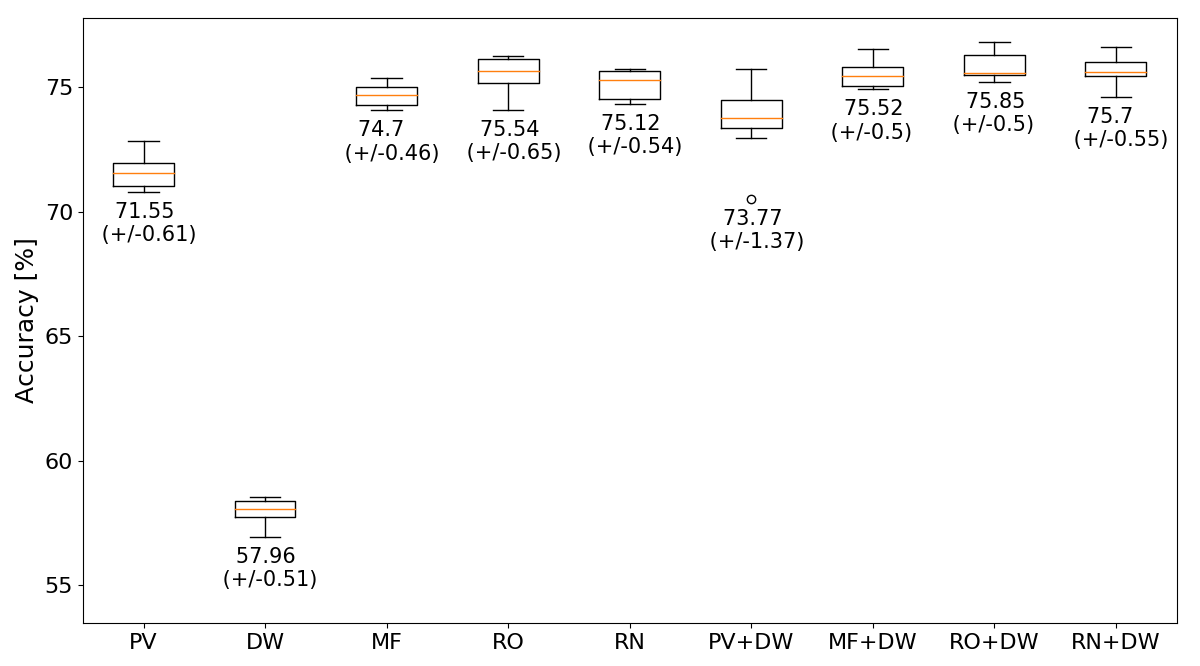}
 \caption{Link Prediction for Genres}\label{fig:genres}
\end{figure}

The link prediction problem is typically defined on graphs where the goal is to predict links that are missing or likely to be created in the future like a probable friendship relation in a social network~\cite{liben2007link}.
In our case, we consider the link prediction task for a specific relation.
We trained our embeddings without considering the respective relations.
Afterward, we take a portion of word pairs which have the relation and a portion of words where no relation exists and train a neural network to predict whether the relation is present.
This is a similar procedure as done in~\cite{lengerich-etal-2018-retrofitting}.\\
\textbf{Network Architecture:} We used an ANN which gets an edge encoded by a source and a target embedding as input.
As shown in Figure~\ref{fig:ann-lp} both embedding layers are processed by an inner layer then get subtracted and the result is processed by another layer which is then connected to the output layer.
The output layer consists only of one output neuron which is translated into a binary value to decide whether the edge is classified as present or not.
Also in this setup, we normalize the input vectors.\\
\textbf{Prediction of Genres for Movies:}
On the movie dataset, we decided to predict the movie-genre relations.
There are $20$ genres in total.
Usually, a movie is assigned to multiple genres.
This prediction is difficult because the remaining metadata provides only limited information about the genres.
Moreover, it is quite subjective in which genre a movie fits and it can be assumed that the genre information is incomplete since some movies are not assigned to a genre.
Furthermore, the genre information is rather diverse.
An example of a genre is ``TV Movie'' which refers to the media it is published in while the majority of the genres refer to content aspects like this is the case for ``Horror'' or ``Comedy''.
It can, therefore, be expected that the accuracy is quite low.
We trained our neural network on sets of $5,000$ movie-genre relations and $5,000$ arbitrary connections between movies and genres which are not in the dataset and serve as negative examples.
We repeated this $10$ times.
The results are presented in Figure~\ref{fig:genres}.
The DeepWalk (DW) embeddings usually used for link prediction fail in this setting.
This is probably the case because the node vector of the genres are quite similar since all of those nodes have only a single edge to the same blank node.
The retrofitted vectors clearly outperform the plain word embeddings.
The standard retrofitting approach (MF) is slightly outperformed by the relational retrofitting approaches.
In combination with node embeddings, text-based approaches achieve better results.
\begin{flushright}

\end{flushright}
\section{Related Work}
\label{sec:relatedwork}
In-database ML is getting more and more important and therefore reflects a very active research field \cite{faerber2012, feng2012, hellerstein2012, kraska2013, cai2013, kara2018}.
However, to the best of our knowledge there exists no work on how to learn a representation that incorporates textual information together with knowledge from a relational database system that can be used to train and serve ML models.\\
\textbf{Representation of Text Values:}
While there is not much related work on the topic of text value representations in database systems for ML, there are some recent papers for spreadsheet data and other semi-structured data sources.
In~\cite{Bordawekar:2017:UWE:3076246.3076251} word embeddings are trained on text serialized from the database system.
In this way, the textual information is combined with the relational information of the database.
However, additional knowledge from external text sources is not considered.
Moreover, word embedding techniques like the Skip-Gram model~\cite{NIPS2013_5021} need large amounts of text values to create good representations which capture the semantic of word accurately.
Since pre-trained word embeddings capture the semantic similarity of words in the natural language, they can be used as a numerical representation for text values.
This was done in Deeper~\cite{ebraheem2018distributed}, an approach for entity resolution on tabular data.
Here, text values are first represented by word embeddings and processed row-wise by an LSTM network that generates so-called tuple embeddings which are used for entity resolution.
Text values which have no word embedding representation are treated by incorporating columnar and row-wise relations.
Deeper only processes single tables and can not consider foreign key relations.
The entity linking system IDEL~\cite{kilias2018idel} also created tuple embeddings based on the concatenation of feature vectors for attributes and foreign keys.
Text attributes are encoded using neural embeddings.
In DataWig~\cite{biessmann2018deep}, text values are represented by n-gram hashing.
In~\cite{fernandez2019termite} a system is proposed which is able to combine heterogeneous data sources based on an embedding approach.\\
\textbf{Retrofitting Approaches:}
We used a retrofitting-based model for our vector representation.
There are several papers proposing different adaptations of the model proposed by Faruqui et al.~\cite{faruqui2014retrofitting} for specific tasks and data sources:
In \cite{kiela2015specializing} the authors propose a model that specializes a word embedding to either capture similarity or relatedness.
The disambiguation between similar and dissimilar relations was investigated in \cite{mrkvsic2016counter}.
Functional relations of property graphs can be retrofitted as proposed by~\cite{lengerich-etal-2018-retrofitting}.
While Faruqui et al. propose to update the vectors one by one, we used a matrix formulation which updates whole vectors at once.
A similar matrix update for retrofitting was first proposed by~\cite{speer2016ensemble}.\\
\textbf{Text-Relation Joined Embeddings:}
Alternatively, vector representations can be learned by models which consider relational information in the form of property graphs and word co-oc\-cur\-rences in one objective function.
Thus, instead of incorporating the relational information in a post-processing step, it is directly considered during the processing of the text.
However, these joined learning approaches~\cite{wang2014knowledge, xu2014rc, yu2014improving, zhong2015aligning} can hardly be adapted to the relations of text values in relational database systems.
On one hand, syntactical equivalent text values that occur in different tables can carry a different semantic.
On the other hand, a text value in a database system is not equivalent to a token in a text.
Moreover, large text corpora of domain-specific text might not be available.
According to the experiments in~\cite{faruqui2014retrofitting} retrofitting can improve word embeddings according to intrinsic tasks better than the joined learning approach proposed by \cite{yu2014improving} which also favors a post-processing approach.
\section{Conclusion}
\label{sec:conclusion}
In this paper, we presented a novel approach for generating learned representations for text values in database systems which find application in a wide range of ML tasks.
We introduced \emph{RETRO}, a novel framework which allows generating relational, retrofitted embeddings for an arbitrary database in a fully automated fashion.
In this way, one can easily apply machine learning tasks like regression, binary, and multi-class classification tasks to databases.
We validated \emph{RETRO} experimentally by building standard feed-forward ANNs for classification and regression tasks.
The experiments exhibit that \emph{RETRO} embeddings outperform pre-trained, plain embeddings but also retrofitted embeddings learned using the approach of Faruqui et al.
In addition, we showed the applicability of the \emph{RETRO}-generated embeddings for data integration tasks like missing value imputation and link prediction where it achieves state-of-the-art performance.
%
\bibliographystyle{abbrv}
\bibliography{references}

\begin{thebibliography}{10}

\bibitem{alghunaim2015vector}
A.~Alghunaim, M.~Mohtarami, S.~Cyphers, and J.~Glass.
\newblock {A} {V}ector {S}pace {A}pproach for {A}spect {B}ased {S}entiment
  {A}nalysis.
\newblock In {\em Proceedings of the 1st Workshop on Vector Space Modeling for
  Natural Language Processing}, pages 116--122, 2015.

\bibitem{biessmann2018deep}
F.~Biessmann, D.~Salinas, S.~Schelter, P.~Schmidt, and D.~Lange.
\newblock {D}eep {L}earning for {M}issing {V}alue {I}mputation in {T}ables with
  {N}on-{N}umerical {D}ata.
\newblock In {\em Proceedings of the 27th ACM International Conference on
  Information and Knowledge Management}, pages 2017--2025. ACM, 2018.

\bibitem{bojanowski2017enriching}
P.~Bojanowski, E.~Grave, A.~Joulin, and T.~Mikolov.
\newblock {E}nriching {W}ord {V}ectors with {S}ubword {I}nformation.
\newblock {\em Transactions of the Association for Computational Linguistics},
  5:135--146, 2017.

\bibitem{Bordawekar:2017:UWE:3076246.3076251}
R.~Bordawekar and O.~Shmueli.
\newblock {U}sing {W}ord {E}mbedding to {E}nable {S}emantic {Q}ueries in
  {R}elational {D}atabases.
\newblock In {\em Proceedings of the 1st Workshop on Data Management for
  End-to-End Machine Learning}, DEEM'17, pages 5:1--5:4, New York, NY, USA,
  2017. ACM.

\bibitem{cai2018comprehensive}
H.~Cai, V.~W. Zheng, and K.~C.-C. Chang.
\newblock {A} {C}omprehensive {S}urvey of {G}raph {E}mbedding: {P}roblems,
  {T}echniques, and {A}pplications.
\newblock {\em IEEE Transactions on Knowledge and Data Engineering},
  30(9):1616--1637, 2018.

\bibitem{cai2013}
Z.~Cai, Z.~Vagena, L.~Perez, S.~Arumugam, P.~J. Haas, and C.~Jermaine.
\newblock {S}imulation of {D}atabase-{V}alued {M}arkov {C}hains {U}sing
  {S}im{SQL}.
\newblock In {\em Proceedings of the 2013 ACM SIGMOD International Conference
  on Management of Data}, SIGMOD '13, pages 637--648, New York, NY, USA, 2013.
  ACM.

\bibitem{dozat2016incorporating}
T.~Dozat.
\newblock {I}ncorporating {N}esterov {M}omentum into {A}dam.
\newblock In {\em ICLR Workshop, (1):2013–2016}, 2016.

\bibitem{ebraheem2018distributed}
M.~Ebraheem, S.~Thirumuruganathan, S.~Joty, M.~Ouzzani, and N.~Tang.
\newblock {D}istributed {R}epresentations of {T}uples for {E}ntity
  {R}esolution.
\newblock {\em Proceedings of the VLDB Endowment}, 11(11):1454--1467, 2018.

\bibitem{faerber2012}
F.~F{\"a}rber, N.~May, W.~Lehner, P.~Gro{\ss}e, I.~M{\"u}ller, H.~Rauhe, and
  J.~Dees.
\newblock The {SAP} {HANA} {D}atabase - {A}n {A}rchitecture {O}verview.
\newblock {\em IEEE Data Eng. Bull.}, 35:28--33, 03 2012.

\bibitem{faruqui2014retrofitting}
M.~Faruqui, J.~Dodge, S.~K. Jauhar, C.~Dyer, E.~Hovy, and N.~A. Smith.
\newblock {R}etrofitting {W}ord {V}ectors to {S}emantic {L}exicons.
\newblock In {\em Proceedings of the 2015 Conference of the North American
  Chapter of the Association for Computational Linguistics: Human Language
  Technologies}, pages 1606--1615, 2015.

\bibitem{faruqui2015sparse}
M.~Faruqui, Y.~Tsvetkov, D.~Yogatama, C.~Dyer, and N.~A. Smith.
\newblock {S}parse {O}vercomplete {W}ord {V}ector {R}epresentations.
\newblock In {\em Proceedings of the 53rd Annual Meeting of the Association for
  Computational Linguistics and the 7th International Joint Conference on
  Natural Language Processing (Volume 1: Long Papers)}, pages 1491--1500, 2015.

\bibitem{feng2012}
X.~Feng, A.~Kumar, B.~Recht, and C.~R{\'e}.
\newblock {T}owards a {U}nified {A}rchitecture for in-{RDBMS} analytics.
\newblock In {\em Proceedings of the 2012 ACM SIGMOD International Conference
  on Management of Data}, SIGMOD '12, pages 325--336, New York, NY, USA, 2012.
  ACM.

\bibitem{fernandez2019termite}
R.~C. Fernandez and S.~Madden.
\newblock {T}ermite: {A} {S}ystem for {T}unneling {T}hrough {H}eterogeneous
  {D}ata.
\newblock In {\em Proceedings of the Second International Workshop on
  Exploiting Artificial Intelligence Techniques for Data Management}, page~7.
  ACM, 2019.

\bibitem{goikoetxea2016single}
J.~Goikoetxea, E.~Agirre, and A.~Soroa.
\newblock {S}ingle or {M}ultiple? {C}ombining {W}ord {R}epresentations
  {I}ndependently {L}earned from {T}ext and {W}ord{N}et.
\newblock In {\em Thirtieth AAAI Conference on Artificial Intelligence}, 2016.

\bibitem{goyal2018graph}
P.~Goyal and E.~Ferrara.
\newblock {G}raph {E}mbedding {T}echniques, {A}pplications, and {P}erformance:
  {A} {S}urvey.
\newblock {\em Knowledge-Based Systems}, 151:78--94, 2018.

\bibitem{gunther2018freddy}
M.~G{\"u}nther.
\newblock {FREDDY}: {F}ast {W}ord {E}mbeddings in {D}atabase {S}ystems.
\newblock In {\em Proceedings of the 2018 International Conference on
  Management of Data}, pages 1817--1819. ACM, 2018.

\bibitem{hellerstein2012}
J.~M. Hellerstein, C.~R{\'e}, F.~Schoppmann, D.~Z. Wang, E.~Fratkin,
  A.~Gorajek, K.~S. Ng, C.~Welton, X.~Feng, K.~Li, and A.~Kumar.
\newblock {T}he {MAD}lib {A}nalytics {L}ibrary: or {MAD} {S}kills, the {SQL}.
\newblock {\em Proceedings of the VLDB Endowment}, 5(12):1700--1711, Aug. 2012.

\bibitem{iwata2018unsupervised}
T.~Iwata and N.~Ueda.
\newblock {U}nsupervised {O}bject {M}atching for {R}elational {D}ata.
\newblock {\em arXiv preprint arXiv:1810.03770}, 2018.

\bibitem{DBLP:journals/corr/JastrzebskiLC17}
S.~Jastrzebski, D.~Lesniak, and W.~M. Czarnecki.
\newblock {H}ow to evaluate word embeddings? {O}n importance of data efficiency
  and simple supervised tasks.
\newblock {\em CoRR}, abs/1702.02170, 2017.

\bibitem{kara2018}
K.~Kara, K.~Eguro, C.~Zhang, and G.~Alonso.
\newblock {C}olumn{ML}: {C}olumn-{S}tore {M}achine {L}earning with
  {O}n-{T}he-{F}ly {D}ata {T}ransformation.
\newblock {\em Proceedings of the VLDB Endowment}, 12:348--361, 12 2018.

\bibitem{kiela2015specializing}
D.~Kiela, F.~Hill, and S.~Clark.
\newblock {S}pecializing {W}ord {E}mbeddings for {S}imilarity or {R}elatedness.
\newblock In {\em Proceedings of the 2015 Conference on Empirical Methods in
  Natural Language Processing}, pages 2044--2048, 2015.

\bibitem{kilias2018idel}
T.~Kilias, A.~L{\"o}ser, F.~A. Gers, R.~Koopmanschap, Y.~Zhang, and M.~Kersten.
\newblock {IDEL}: {I}n-{D}atabase {E}ntity {L}inking with {N}eural
  {E}mbeddings.
\newblock {\em arXiv preprint arXiv:1803.04884}, 2018.

\bibitem{kraska2013}
T.~Kraska, A.~Talwalkar, and J.~Duchi.
\newblock {ML}base: {A} {D}istributed {M}achine-learning {S}ystem.
\newblock In {\em In CIDR}, 2013.

\bibitem{lenci2018}
A.~Lenci.
\newblock {D}istributional {M}odels of {W}ord {M}eaning.
\newblock {\em Annual review of Linguistics}, 4:151--171, 2018.

\bibitem{lengerich-etal-2018-retrofitting}
B.~Lengerich, A.~Maas, and C.~Potts.
\newblock {R}etrofitting {D}istributional {E}mbeddings to {K}nowledge {G}raphs
  with {F}unctional {R}elations.
\newblock In {\em Proceedings of the 27th International Conference on
  Computational Linguistics}, pages 2423--2436, Santa Fe, New Mexico, USA, Aug.
  2018. Association for Computational Linguistics.

\bibitem{liben2007link}
D.~Liben-Nowell and J.~Kleinberg.
\newblock {T}he {L}ink-{P}rediction {P}roblem for {S}ocial {N}etworks.
\newblock {\em Journal of the American society for information science and
  technology}, 58(7):1019--1031, 2007.

\bibitem{NIPS2013_5021}
T.~Mikolov, I.~Sutskever, K.~Chen, G.~S. Corrado, and J.~Dean.
\newblock {D}istributed {R}epresentations of {W}ords and {P}hrases and their
  {C}ompositionality.
\newblock In {\em Advances in Neural Information Processing Systems 26}, pages
  3111--3119. Curran Associates, Inc., 2013.

\bibitem{mrkvsic2016counter}
N.~Mrk{\v{s}}i{\'c}, D.~{\'O}. S{\'e}aghdha, B.~Thomson, M.~Ga{\v{s}}i{\'c},
  L.~M. Rojas-Barahona, P.-H. Su, D.~Vandyke, T.-H. Wen, and S.~Young.
\newblock {C}ounter-fitting {W}ord {V}ectors to {L}inguistic {C}onstraints.
\newblock In {\em Proceedings of the 2016 Conference of the North American
  Chapter of the Association for Computational Linguistics: Human Language
  Technologies}, pages 142--148, 2016.

\bibitem{pennington2014glove}
J.~Pennington, R.~Socher, and C.~D. Manning.
\newblock {G}lo{V}e: {G}lobal {V}ectors for {W}ord {R}epresentation.
\newblock In {\em Empirical Methods in Natural Language Processing (EMNLP)},
  pages 1532--1543, 2014.

\bibitem{perozzi2014deepwalk}
B.~Perozzi, R.~Al-Rfou, and S.~Skiena.
\newblock {D}eepwalk: {O}nline {L}earning of {S}ocial {R}epresentations.
\newblock In {\em Proceedings of the 20th ACM SIGKDD international conference
  on Knowledge discovery and data mining}, pages 701--710. ACM, 2014.

\bibitem{schnabel2015evaluation}
T.~Schnabel, I.~Labutov, D.~Mimno, and T.~Joachims.
\newblock Evaluation methods for unsupervised word embeddings.
\newblock In {\em Proceedings of the 2015 Conference on Empirical Methods in
  Natural Language Processing}, pages 298--307, 2015.

\bibitem{speer2016ensemble}
R.~Speer and J.~Chin.
\newblock {A}n {E}nsemble {M}ethod to {P}roduce {H}igh-{Q}uality {W}ord
  {E}mbeddings.
\newblock {\em arXiv preprint arXiv:1604.01692}, 2016.

\bibitem{srivastava2014dropout}
N.~Srivastava, G.~Hinton, A.~Krizhevsky, I.~Sutskever, and R.~Salakhutdinov.
\newblock {D}ropout: {A} {S}imple {W}ay to {P}revent {N}eural {N}etworks from
  {O}verfitting.
\newblock {\em The journal of machine learning research}, 15(1):1929--1958,
  2014.

\bibitem{vrandevcic2014wikidata}
D.~Vrande{\v{c}}i{\'c} and M.~Kr{\"o}tzsch.
\newblock {W}ikidata: {A} {F}ree {C}ollaborative {K}nowledge {B}ase.
\newblock {\em Communications of the ACM}, 57(10):78--85, 2014.

\bibitem{wang2014knowledge}
Z.~Wang, J.~Zhang, J.~Feng, and Z.~Chen.
\newblock {K}nowledge {G}raph and {T}ext {J}ointly {E}mbedding.
\newblock In {\em Proceedings of the 2014 conference on empirical methods in
  natural language processing (EMNLP)}, pages 1591--1601, 2014.

\bibitem{xu2014rc}
C.~Xu, Y.~Bai, J.~Bian, B.~Gao, G.~Wang, X.~Liu, and T.-Y. Liu.
\newblock {RC}-{NET}: {A} {G}eneral {F}ramework for {I}ncorporating {K}nowledge
  into {W}ord {R}epresentations.
\newblock In {\em Proceedings of the 23rd ACM international conference on
  conference on information and knowledge management}, pages 1219--1228. ACM,
  2014.

\bibitem{yu2014improving}
M.~Yu and M.~Dredze.
\newblock {I}mproving {L}exical {E}mbeddings with {S}emantic {K}nowledge.
\newblock In {\em Proceedings of the 52nd Annual Meeting of the Association for
  Computational Linguistics (Volume 2: Short Papers)}, volume~2, pages
  545--550, 2014.

\bibitem{zhong2015aligning}
H.~Zhong, J.~Zhang, Z.~Wang, H.~Wan, and Z.~Chen.
\newblock {A}ligning {K}nowledge and {T}ext {E}mbeddings by {E}ntity
  {D}escriptions.
\newblock In {\em Proceedings of the 2015 Conference on Empirical Methods in
  Natural Language Processing}, pages 267--272, 2015.

\bibitem{zhou2015representation}
X.~Zhou, X.~Wan, and J.~Xiao.
\newblock {R}epresentation {L}earning for {A}spect {C}ategory {D}etection in
  {O}nline {R}eviews.
\newblock In {\em Twenty-Ninth AAAI Conference on Artificial Intelligence},
  2015.

\end{thebibliography}

\begin{appendix}
\balance
\section{Appendix}
\label{ap:convex}
In the following, we want to prove the convexity of $\Psi(W)$.
Instead of processing $W$ row-wise as this was done in the sections before, we consider each of its elements $w_{i,j}$ separately in function $\Psi$.
This is possible since the quadratic Euclidean distances can be split in a sum of quadratic differences of coordinate values.
\begin{flalign}
	\nonumber
	\Psi(W) =& \sum_{d = 1}^D\sum_{i = 1}^n \Big[ \alpha_i (w_{i,d} - w_{i,d}')^2 + \beta_j \Psi_C(w_{i,d}, W) && \\
	& + \Psi_R(w_{i,d}, W) \Big]
\end{flalign}
$\Psi(W)$ can be split in $\Psi(W) = \hat\Psi(W) + \Psi_\beta(W) + \Psi_\gamma(W)$:
\begin{flalign}
	\nonumber
\hat\Psi(W) = & \sum_{d = 1}^D\sum_{i = 1}^n \Big[ \alpha_i (w_{i,d} - w_{i,d}')^2 \\
\nonumber
 & - \sum_{r \in R}\Big[ \sum_{k:(i,k)\in \widetilde{E_r}} \delta_i^r( w_{i,d} - w_{k,d})^2 \Big]\Big]  &&\\
 \nonumber
\Psi_\beta(W) = & \sum_{d = 1}^D\sum_{i = 1}^n \Big(w_{i,d} - \frac{\sum\limits_{j \in C(i)}{w_{j,d}'}}{|C(i)|}\Big)^2 &&\\
\Psi_\gamma(W) = &  \sum_{d = 1}^D\sum_{i = 1}^n\sum_{r \in R} \Big[ \sum_{j:(i,j)\in E_r} \gamma_i^r ( w_{i,d} - w_{j,d} )^2 \Big] &&
\end{flalign}
We utilize the fact, that a sum of convex functions is also convex.
It is easy to see that $\Psi_\beta(W)$ and $\Psi_\gamma(W)$ are convex functions if all $\gamma_i^r$ and $\beta_i$ values are positive.
$\Psi_\beta(W)$ is a simple $D$-dimensional quadratic function and thus convex.
$\Psi_\gamma(W)$ consists of sums of squared differences which are themselves convex functions.
Hence, in order to prove the convexity of $\Psi(W)$, it is sufficient to prove convexity for $\hat\Psi$.
As a next step, we create the Hessian matrix of all second partial derivatives:
\begin{flalign}
	\nonumber
H  = \begin{pmatrix}
\frac{\partial^2\hat\Psi(W)}{\partial w_{1,1}^2} & \dots & \frac{\partial^2\hat\Psi(W)}{\partial w_{1,1},\partial w_{n,D}} \\
\vdots & \ddots & \vdots\\
\frac{\partial^2\hat\Psi(W)}{\partial w_{n,D}, \partial w_{1,1}} & \dots  & \frac{\partial^2\hat\Psi(W)}{\partial w_{n,D}^2}
\end{pmatrix}\\
\end{flalign}
\begin{flalign}
\nonumber
\frac{\partial^2\hat\Psi(V)}{\partial w_{i,d}^2} =& 2 \Big(\alpha -\sum_{r \in R} \Big[\; \sum_{\mathclap{\substack{k:(i,k) \\ \in \widetilde{E_r}}}} (\delta_i^r+\delta_k^{\bar{r}}) \Big]\Big)\\
\nonumber
\frac{\partial^2\hat\Psi(V)}{\partial w_{i,d}, \partial w_{k,d}} =& \sum_{r \in R}\phi(i,k,r)\;\;\mathit{if}\; i \neq k \\
\nonumber
\phi(i,k,r) =& \begin{cases} 4\delta_i^r & (i,k) \in \widetilde{E_r} \\
 0 & \mathit{otherwise} \\
\end{cases}\\
\frac{\partial^2\hat\Psi(V)}{\partial w_{i,d}, \partial w_{j,d'}} =& 0 \;\;\mathit{if}\;\; d \neq d'
\end{flalign}
$\hat\Psi(W)$ is convex if and only if the Hessian matrix is positive semi-definite.
According to the parameter configuration defined in Section~\ref{subsec:paramaterconf}, the following condition holds:
\begin{flalign}
(i,k) \in \widetilde{E_r} \implies (k,i) \in \widetilde{E_{\bar{r}}} \land \delta_i^r = \delta_k^{\bar{r}} 
\end{flalign}
This leads to the following equivalence:
\begin{flalign}
\frac{\partial^2\hat\Psi(V)}{\partial w_{i,d}, \partial w_{j,'}} = \frac{\partial^2\hat\Psi(V)}{\partial w_{j,d}, \partial w_{i,d}}
\end{flalign}
Subsequently, $H$ is a symmetric matrix.
An interesting matrix property to consider for a symmetric matrix $(a_{ij})$ is the diagonally dominance.
A matrix is diagonally dominant, if in every row $i$ the magnitude of the element on the diagonal $|a_{ii}|$ is greater or equal than the sum of magnitudes of the remaining elements $\sum_{j\neq i}|a_{ij}|$.
If a matrix is diagonally dominant, it is also positive semi-definite.
Therefore, the following is a sufficient condition to show that $\hat\Psi(W)$ is a convex function.
\begin{flalign}
\nonumber
\forall i \in \{1,\dots n\times D\}: |h_{ij}| \geq \sum_{j\neq i}|h_{ij}|\\
\nonumber
\forall i \in \{1,\dots n\}, d\in\{1,\ldots, D\}: \;\;\;\;\;\;\;\;\;\;\;\;\;\;\;\;\;\;\;\;\;\;\;\;\;\;\;\;\;\\\
\nonumber
\Big|\frac{\partial^2\hat\Psi(W)}{\partial w_{i,d}^2}\Big| \geq \sum_{j\neq i} \Big|\frac{\partial^2\hat\Psi(V)}{\partial w_{i,d}, \partial w_{j,d}}\Big|\\
\nonumber
2 \Big|\alpha_i -\sum_{r \in R} \Big[\; \sum_{\mathclap{\substack{j:(i,j) \\ \in \widetilde{E_r}}}} (\delta_i^r+\delta_j^r) \Big]\Big| \geq \sum_{j\neq i}\Big|\sum_{r \in R}\phi(i,j,r)\Big|\\
\Big|\alpha_i -\sum_{r \in R} \Big[\; \sum_{\mathclap{\substack{j:(i,j) \\ \in \widetilde{E_r}}}} 2\delta_i^r \Big]\Big| \geq \sum_{r \in R}\Big|\sum_{\mathclap{\substack{j:(i,j) \\ \in \widetilde{E_r}}}}2\delta_i^r\Big|
\end{flalign}
If we assume that for every $i$ and $r$ the values $a_i$ and $\delta_i^r$ are positive, the solution of this inequality results in Equation~\eqref{eq:proof:convex}:
\begin{flalign}\label{eq:proof:convex}
	\alpha_i \geq 4\sum_{r \in R}\Big[\;\sum_{\mathclap{\substack{j:(i,j) \\ \in \widetilde{E_r}}}}\delta_i^r\Big]
\end{flalign}
\end{appendix}

%
% ...

\end{document}